\documentclass[]{aa}
\usepackage{graphicx}
\usepackage{subfig}
\usepackage{caption}
\usepackage{float}
\usepackage{threeparttable}
\usepackage{booktabs}
\usepackage{txfonts}
\usepackage{CJKutf8}
\usepackage{amsmath}
\usepackage{placeins}
\usepackage[normalem]{ulem}

\newcommand*\kms{km s$^{-1}$}
\newcommand*\twelve{$^{12}$CO $J=3\to2$ }
\newcommand*\thirteen{$^{13}$CO $J=3\to2$ }
\newcommand*\eighteen{C$^{18}$O $J=3\to2$ }
\newcommand*\Ta{$T_A^*$ }
\newcommand*\Tmb{$T_{MB}$ }
\usepackage{hyperref}
\hypersetup{colorlinks=true,linkcolor=blue,citecolor=blue,filecolor=blue,urlcolor=blue}
\begin{document}
\title{The vertical distribution of molecular gas \\in the fourth quadrant of the Milky Way}
\author{ L.-H. Lin \begin{CJK*}{UTF8}{gbsn}(林良浩)\end{CJK*} \inst{1}
    \and F. Wyrowski \inst{1}
    \and S. Neupane \inst{1} 
    \and D. Colombo \inst{2}
    \and K. M. Menten\inst{1}\thanks{Professor K. M. Menten passed away in December 2024.}
    }
\institute{
   Max-Planck-Institut für Radioastronomie, Auf dem Hügel 69, D-53121 Bonn, Germany \\
        \email{lianghaolin518@gmail.com, wyrowski@mpifr-bonn.mpg.de}
    \and Argelander-Institut f\"ur Astronomie, Universit\"at Bonn, Auf dem H\"ugel 71, D-53121 Bonn, Germany
   }
\date{Received January 23, 2026 / Revised October 25, 2026 / Accepted October 28, 2026}
\abstract
{The vertical distribution of molecular gas in the Milky Way disk is poorly studied, especially in the fourth quadrant and the far side of the Galaxy. While the old surveys conducted more than 20 years ago suffered from coarse beam and non-Nyquist sampling, most of the recent surveys are limited to $|b|<1\degr$ and/or focus on the Northern Galactic plane.}
{We want to examine the molecular gas's vertical distribution and properties in the fourth quadrant in order to check the impact of Galactic structure on current star formation.}
{We mapped two arm tangent regions ($l=[329\degr.0, 327\degr.0]$ for the Norma-Outer arm, and $l=[310\degr.0, 308\degr.5]$ for the Scutum-Centaurus arm) with the \twelve line using the 7-pixel LASMA receiver of the APEX 12 m telescope, covering up to $b=[-2\degr, +2\degr]$ (or $|z|\sim200$ pc near the tangent points). The noise level of the \twelve  line is about 0.4 K [$T_{MB}$] in velocity channels of 0.5 \kms.}
{The vertical distribution around the tangent regions of the spiral arm is fitted with two Gaussian components, the narrow one with an FWHM thickness of 25 pc and the wide one of about 100 pc. 
Two explanations are proposed: they are two sub-components of the thin disk (the narrowest tracing the spiral arm and the other the inter arms components), or they trace both thin and thick disks that show a smaller thickness when observed with CO 3-2 than when observed with CO 1-0. 
The latter indicates that the density distribution of the interstellar medium exhibits a steeper gradient at larger heights above the plane.
Clumps in the two tangent regions are searched and divided into groups according to their relative height.
Compared with the low relative height clumps ($|z|<50 $ pc), the median relative height clumps ($50~\mathrm{pc}<|z|<100~\mathrm{pc}$) have  lower excitation temperature and surface density, with fewer high values in the distribution. A similar tentative trend is found for the mass.} 
{}
\titlerunning{Vertical distribution of molecular gas in Galactic 4th quadrant}
\authorrunning{L.-H. Lin et al.}
\maketitle
\section{Introduction}
“How is the molecular gas distributed with increasing radius and height above the plane and with respect to spiral structure?”
The question about the vertical distribution of the molecular gas disk in the review of  \citet{2015ARA&A..53..583H} remains to be understood.
The Galactic disk is a composite of stars and interstellar medium (ISM) formed primarily of \ion{H}{I} gas and H$_2$ gas.
All of them can be further divided into two components, namely the thin disk and the thick disk.
The vertical distributions of the stellar disk and the \ion{H}{I} disk are better understood than the H$_2$ disk, as we summarize in the following.

\citet{2016ARA&A..54..529B} reviewed the properties of the Milky Way and recommended that the stellar disk has an exponential scale height with a reference value in the solar neighbourhood of $300\pm50$ pc for the thin stellar disk and $900\pm180$ pc for the thick stellar disk, which equals to full-width half-maximum (FWHM) heights of $z^t_*=420\pm70$ pc and $z^T_*=1250\pm250$ pc, respectively.

Recently, \citet{2023ARA&A..61...19M} reviewed the \ion{H}{I} studies in the Milky Way.
For the solar neighbourhood, the \ion{H}{I} gas disk has a thin disk about $z^t_{\ion{H}{I}}=400$ pc FWHM, similar to the stellar disk.
The \ion{H}{I} gas is usually analysed only beyond the solar circle (or even $R_{GC}>10$ kpc) to overcome the ambiguity of kinematic distance, as \ion{H}{I} is more extended than CO and more difficult to link the emission with the spiral arm.

\citet{2015ARA&A..53..583H} reviewed the CO gas studies in the Milky Way. 
Using data observed mainly in the 1980s, the authors reviewed that the molecular gas disk shows a slight rise in the thickness from about 90 pc at Galactocentric radius $R_{GC}=2$ kpc to about $z^t_{\mathrm{CO}}=120$ pc FWHM at the solar neighbourhood ($R_{GC}=8.2$ kpc), with a mean displacement of the mid-plane about 30 pc in the halfway (their Figure 6).
However, these results are mainly reported as functions of $R_{GC}$, which omits the dependence on the azimuth angle $\phi$ and hence ignores the spiral structure of our galaxy.

The old studies listed in \citet{2015ARA&A..53..583H} had used data from \citet{2001ApJ...547..792D} or even older, which have coarse spatial resolution and high noise, and were limited to the first quadrant since the telescopes used in these surveys were mainly located on the Northern hemisphere.
Most of the recent unbiased large-area Galactic-plane surveys of molecular gas are limited to $|b|<1\degr$ or an even narrower belt, which limits the possibility of exploring the vertical distribution of molecular gas using these data sets, e.g. the FCRAO Galactic Ring Survey \citep{2006ApJS..163..145J}, the Mopra Southern Galactic Plane CO Survey \citep{2018PASA...35...29B}, and the APEX SEDIGISM \citep{2017A&A...601A.124S} and OGHReS \citep{2021A&A...655L...2C} surveys of CO. 
On the other hand, dust continuum surveys like the APEX ATLASGAL survey \citep{2009A&A...504..415S} and the Herschel Hi-GAL survey \citep{2010PASP..122..314M} go beyond $|b|<1\degr$ but cannot distinguish multiple spiral-arm components in the line-of-sight.

The only recent systematic study was made by \citet{2021ApJ...910..131S} using the MWISP survey.
The designation of the MWISP survey was described by \citet{2019ApJS..240....9S} in detail.
This study shows that one can still identify thin and thick molecular disks with high-resolution observations, although with a relatively high RMS noise at 0.5 K for 0.16 \kms \ channel rather than the old 0.1 K data of \citet{1994ApJ...436L.173D} in which the thick molecular disk was discovered.
They reported that the FWHM scale heights of the two components are about $z\mathrm{^t_{CO}}=85$ and $z\mathrm{^T_{CO}}=280$ pc, and the thick disk dominates at $|z|>100$ pc, similar to the old discovery result of \citet{1994ApJ...436L.173D}.
In addition, they found that the FWHM and midplane displacement of the thin disk have a variability depending on the longitude of tangent regions.
This MWISP study is limited to the tangent regions in the first quadrant, where the distance can be well determined.
However, due to their 50\arcsec\ resolution, the MWISP study had to include both clouds on the spiral arms and inter-arm clouds for better statistics, but the authors did not compare their high-latitude clouds with the low-latitude sample.
In addition, their samples had a large range of Galactocentric distances, which may include other environmental effects rather than the height of targets.

The 7-pixel LASMA receiver of the Atacama Pathfinder EXperiment (APEX) 12-m telescope gives us an opportunity to explore the vertical distribution of molecular gas in the fourth quadrant of our Milky Way with CO isotopologues $J=3\to2$ transitions and a beam size of about 18 arcsec.
With this higher resolution, we are able to resolve a linear scale of about 0.5 pc at two tangent positions (distances about 5 kpc), which allows us to detect a  statistically significant number of molecular structures and to examine the vertical distribution of their physical parameters for dense structures rather than large clouds for the first time to our knowledge.
Across this work, we call a structure as a clump when it is a hierarchical, connected region of CO emission bounded by an iso-intensity surface and without sub-structure in our our 0.5 pc beam (see Sect. \ref{sect:result:clump} for details).

This paper is arranged as follows. 
In Sect. \ref{sect:observation} we describe our observations and the data used in this study.
The reduction method is described in Sect. \ref{sect:reduction}.
The vertical distribution of molecular gas and the properties of clumps around the tangent regions are derived in Sect. \ref{sect:result:vertical} and \ref{sect:result:clump}.
In Sect. \ref{sect:Discussion}, we compare the integrated properties and clumps' statistics with the old studies.

\section{Observations}
\label{sect:observation}
\begin{figure}[t]
    \centering
    \includegraphics[width=\linewidth]{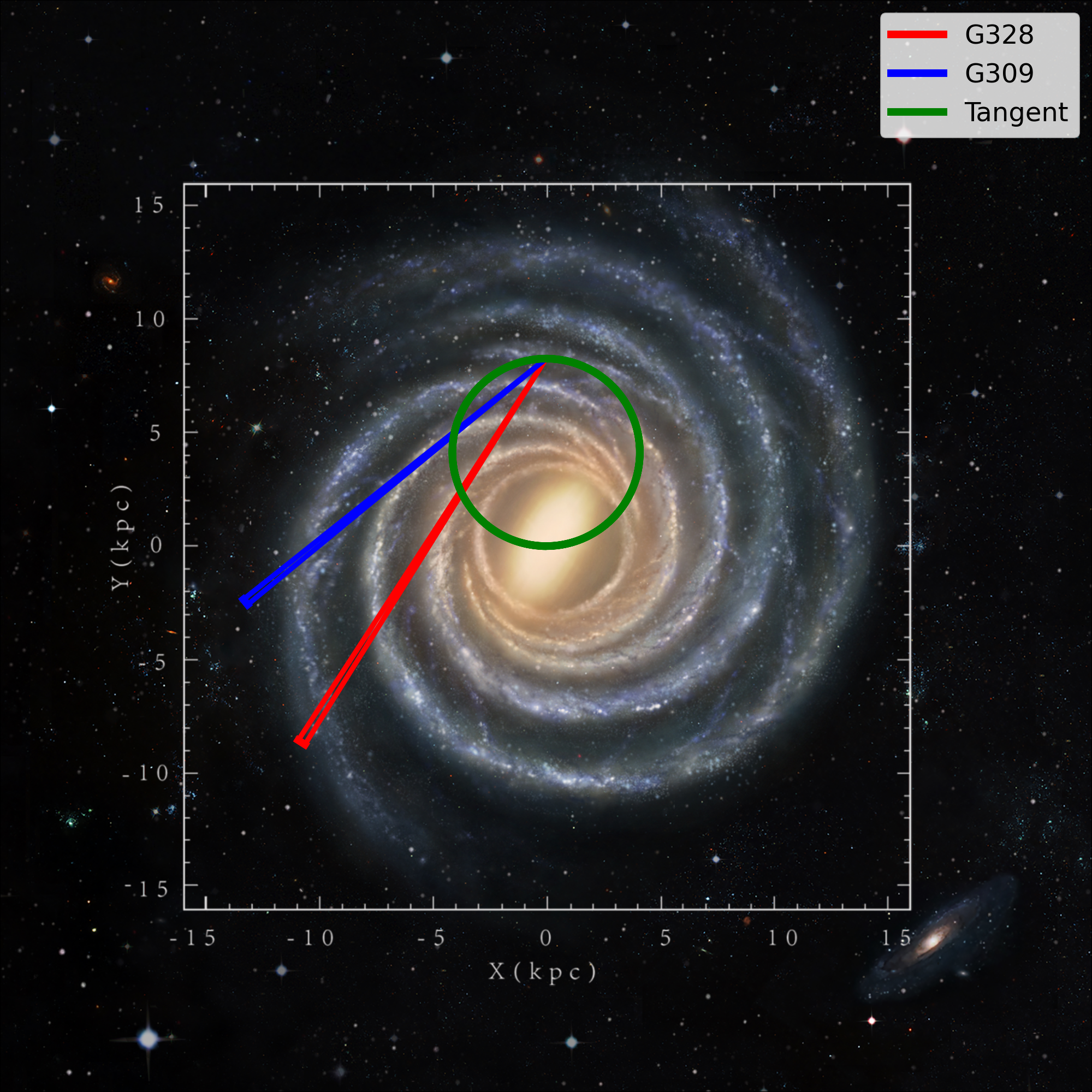}
    \caption{The region we have covered in this work. Red is the G328 region and blue the G309 region. The green circle shows the tangent region, defined as the loci of the maximum radial velocity. The spiral arm tangents are the crossing of the observed line of sight with the green circle. This plot is reproduced using the Xing-Wu Zheng \& Mark Reid (BeSSeL/NJU/CFA) 2019 figure (\url{https://astronomy.nju.edu.cn/xtzl/EN/}). }
    \label{fig:2region}
\end{figure}

We mapped two spiral-arm tangent regions in the fourth quadrant of our Galaxy ($l=[329.0\degr, 327.0\degr]$ for the Norma-Outer arm, and $l=[310.0\degr, 308.5\degr]$ for the Scutum-Centaurus arm) with the $J=3\to2$ lines of CO isotopologues using the LASMA receiver of the APEX 12 m telescope, covering up to $b=[-2.0\degr, +2.0\degr]$ (or $|z|\sim200$ pc near the tangent points).
These two tangents are selected to maximize the amount of material sampled at an equivalent and well-determined distance within each region.
These two regions are shown in Fig. \ref{fig:2region} and their physical parameters are listed in Table \ref{tab:2tangent} with their distances derived from \citet{2019ApJ...885..131R}. 
The three-dimension position $(l, \bar b, d)$ is converted to the cylindrical representation of Galactocentric frame $(R_{GC}, \phi, \bar z)$ with solar system values as $R_{sun}=8.122$ kpc \citep{2018A&A...615L..15G} and $z_{sun}=20.8$ pc \citep{2019MNRAS.482.1417B} using the Galactocentric frame of \textit{Astropy}.
The height value of $z$ across this paper refers to the "true" Galactic mid-plane instead of the $b=0\degr$ plane, considering the sun is above the mid-plane with $z_{sun}=20.8$ pc.
In some other papers, \citep[e.g.][]{2021ApJ...910..131S}, they use the $b=0\degr$ plane as their reference, i.e. $z=d\sin b$, which leads to some difference between regions with different longitudes.
Here, the spiral-arm tangent regions are where the spiral arms cross the tangent region (the green circle in Fig. \ref{fig:2region}, showing the location of the maximum radial velocities).
We note that there is another definition of spiral-arm tangent, which is the tangent of a spiral arm from an observer in the solar system (i.e. maximum longitude).
However, distances and line-of-sight velocities of regions under this definition are not well-determined.

\begin{table*}[htbp!]
\centering
\caption{Parameters of the two spiral-arm tangent regions. }
\label{tab:2tangent}
\begin{tabular}{cccccr}
\toprule
Velocity range          & Velocity              & Distance              &       Radius   &   Azimuth    & $b=0$ Offset \\ 
$v_{LSR}$ (km s$^{-1}$) & $\bar{v}$ (km s$^{-1}$)& $d\pm\sigma_d$ (kpc) & $R_{GC}$ (kpc) & $\phi$ (deg) & $z_{b=0}$ (pc)  \\
\midrule
\multicolumn{6}{c}{G328 Norma tangent region} \\
$[-120,-100]$ & $-105$ & $ 6.1\pm0.7$ &  4.4 & $-132$ & $+7.6$ \\
\midrule
\multicolumn{6}{c}{G309 Centaurus tangent region}\\
$[-70,-57]$ & $ -60$ & $ 5.1\pm0.8$ &  6.3 & $-141$ & $+11.0$ \\
\bottomrule
\end{tabular}
\end{table*}

Our survey regions were divided into many $30'\times30'$ On-the-fly (OTF) fields.
The step size between two scan lines for the centre pixel of the LASMA array receiver was carefully chosen as 47.1 arcsec to minimise the overlap among the scan lines of all pixels while still ensuring Nyquist sampling.
To ensure full coverage of the fields, the OTF region was set to $1800"\times1837"$, hence we had overlapping edges of the fields.

The Norma-Outer arm tangent region around $l=328\degr$ (G328 region hereafter) was mapped from June to August 2023.
During that time, the LASMA receiver provided an IF bandwidth of 4$\to$8 GHz.
We covered the \twelve and \thirteen and CS $J=7\to6$ lines with an LO setup at 338.3 GHz.
The Scutum-Centaurus arm tangent region around $l=309\degr$ (G309 region hereafter) was mapped from October to December 2023 after the IF bandwidth of the receiver had been extended to 4$\to$10 GHz. 
Hence, we additionally covered the \eighteen line with a slightly different LO setup at 338.7 GHz for the latter region.

Each field was mapped by OTF scans in both X and Y scan directions. 
One coverage took about 40 minutes of telescope time, and it took 7 hours to finish a square degree.
For the G328 region, we got additional time to repeat all fields with one more coverage (two more coverages in some noisy fields) to get a homogeneous noise distribution.

\section{Data Reduction}
\label{sect:reduction}
We developed a scan-based pixel efficiency model to correct the antenna temperature $T_A^*$ observed by the 7-pixel array receiver into the main beam brightness temperature $T_{MB}$ scale, considering the rotation of the LASMA receiver (see Appendix \ref{app:eff_model}).
The full-width half-maximum (FWHM) beam size was about 18.1 arcsec.
The main beam efficiency was $\eta_{MB}=0.76\pm0.06$ as derived from Jupiter observations during the observation period, and the forward efficiency was 0.95 \footnote{\url{https://www.apex-telescope.org/telescope/efficiency/index.php}}.

To facilitate the comparison of different lines and increase the signal-to-noise ratio, data cubes for all the lines were resampled to the same velocity channel of 0.5 km s$^{-1}$ and gridded to the same pixel size of $9\arcsec$.
The data reduction was conducted using the GILDAS-CLASS \footnote{\url{http://www.iram.fr/IRAMFR/GILDAS/}} package.

\begin{figure}[htbp!]
    \centering
    \includegraphics[width=1\linewidth]{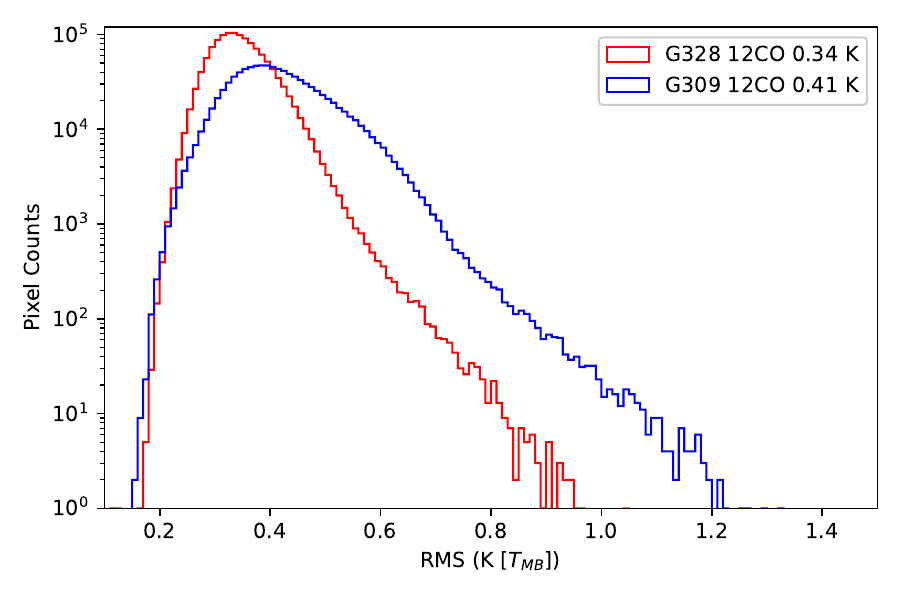}
    \caption{The distribution of the RMS noise in each spatial pixels of the \twelve line for the two regions, in $T_{MB}$ scale. The median values for the two regions are 0.34 K for the G328 region and 0.41 K for the G309 region.}
    \label{fig:rms-distribution}
\end{figure}

We derive the RMS noise of the \twelve line in each spatial pixel of the data cube (Fig. \ref{fig:rms-maps}, and show its statistical distribution in Fig. \ref{fig:rms-distribution}. The median noises of our \twelve cubes are 0.34 K for the G328 region and 0.41 K for the G309 region in $T_{MB}$ scale.
Due to the atmosphere water line at 325 GHz, the \thirteen and \eighteen lines are about 50\% and 80\% noisier than \twelve.
The CS $J=5\to4$ line is only detected towards the centres of a few clumps.
Given the sensitivity of the observations, we will focus on the \twelve results in this work because it is the most extended line with the lowest level of noise. The \twelve line also traces a less dense medium with respect to the other two transitions \citep{2015ARA&A..53..583H}.

\section{Result}
\label{sect:result}

\subsection{Vertical distribution of gas}
\label{sect:result:vertical}

\begin{figure}[tbp!]
    \subfloat[]{
        \begin{minipage}{1\linewidth}
        \centering
        \includegraphics[width=1\textwidth]{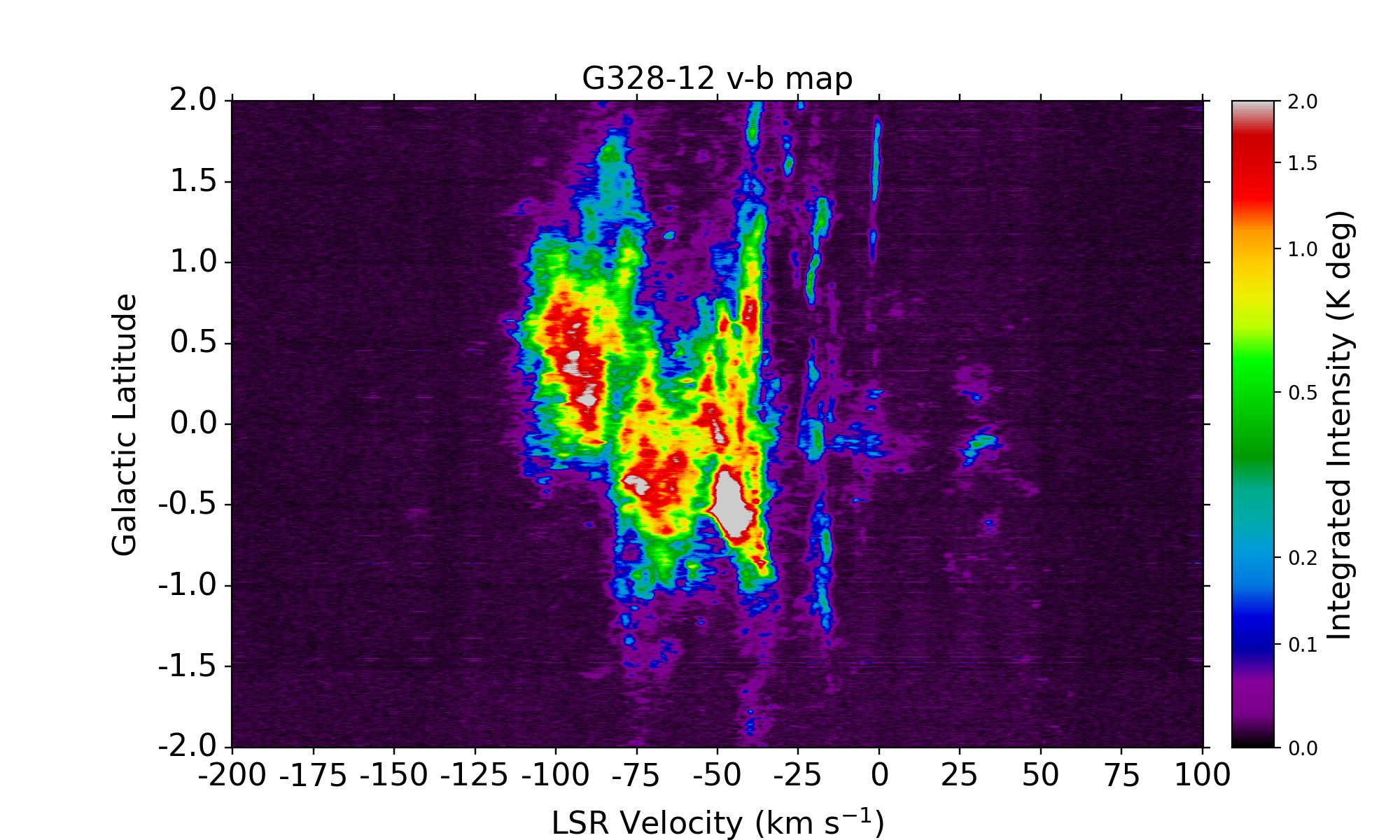}
        \end{minipage} } 
        
    \subfloat[]{
        \begin{minipage}{1\linewidth}
        \centering
        \includegraphics[width=1\textwidth]{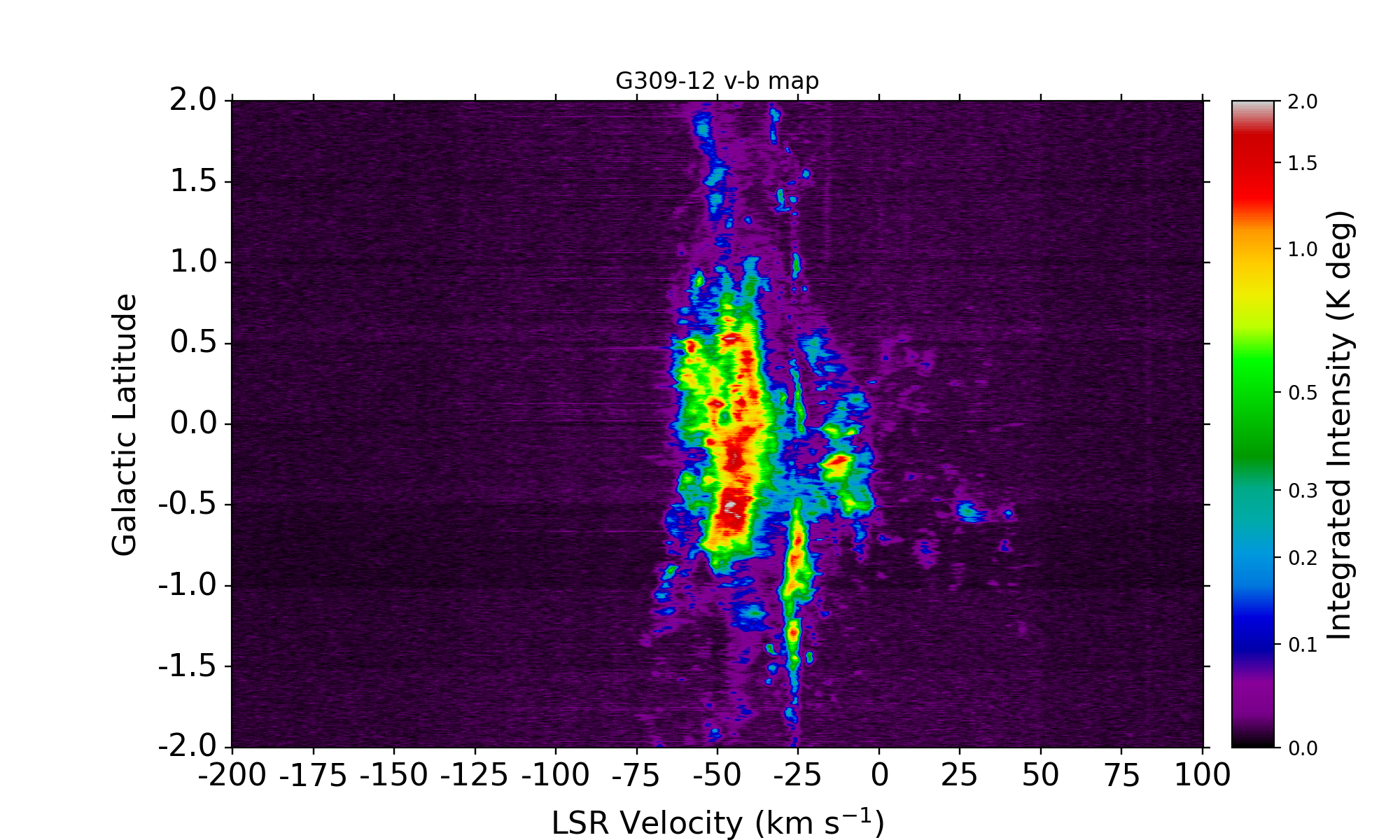}
        \end{minipage} } 
    \caption{The \twelve velocity-latitude maps of the two regions.}
    \label{fig:vbmap}
\end{figure}

The main goal of this work is to study the vertical distribution of molecular gas in the tangent region. 
Here we report the results of the integrated vertical distribution of detected gas.
The velocity-latitude maps (v-b maps) of the two regions are shown in Fig. \ref{fig:vbmap}.
We defined the inner velocity boundary of the tangent region as the half peak velocity in the average spectrum of the specific spiral arm ($-100$ and $-57$ \kms, respectively), and the outer velocity boundary is a noise truncation.
The integrated intensity maps for the tangent region and other individual spiral arms of these two regions are shown in Fig. \ref{fig:m0maps-G328} and \ref{fig:m0maps-G309}, respectively.

\begin{figure}[htbp!]
    \subfloat[]{
        \begin{minipage}{0.95\linewidth}
        \centering
        \includegraphics[width=\textwidth]{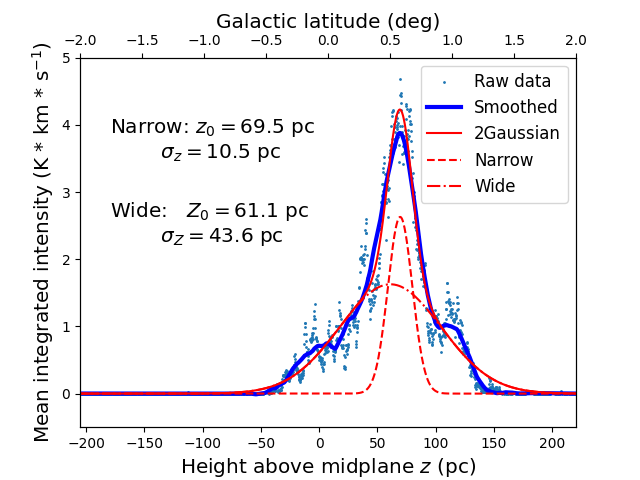}
        \end{minipage} } \\
        
    \subfloat[]{
        \begin{minipage}{0.95\linewidth}
        \centering
        \includegraphics[width=\textwidth]{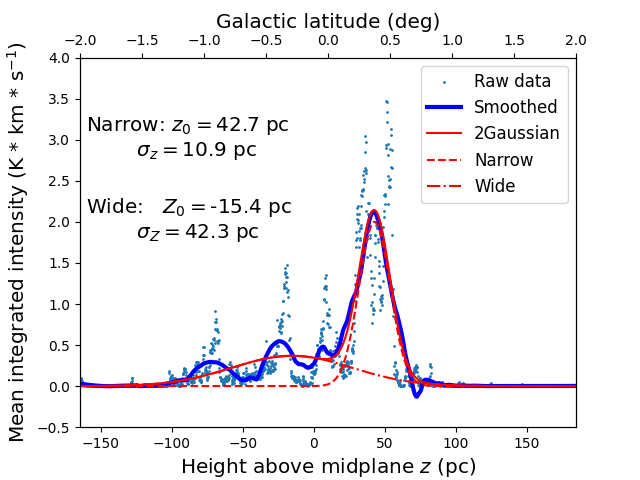}
        \end{minipage} } 
    \caption{The vertical distribution of \twelve emission around the tangent velocity of the (a) G328 and (b) G309 region. These two plots are the fourth quadrant counterparts of Figure 4 of \citet{2021ApJ...910..131S}.}
    \label{fig:2gaussian}
\end{figure}

The vertical distribution of total molecular gas is estimated by the intensity integrated over the longitude and velocity. We show the vertical distribution of the tangent regions and fit them with two Gaussian components, as shown in Fig. \ref{fig:2gaussian}.
The observed distribution can be fitted with two Gaussians, one narrow and one wide. The best fit for the narrow components has a  thickness of about 11 pc ($1\sigma$) or in FWHM 25 pc, and the best fit for the thickness of the wide components is about 43 pc ($1\sigma$) or 100 pc (FWHM).
We discuss these two-component vertical distribution further in Sect. \ref{sect:Discussion:intergrate}.

\subsection{Clumps in the tangent regions}
\label{sect:result:clump}
The \twelve clumps in the tangent regions are searched using the dendrogram method \citep{2008ApJ...679.1338R}.
The settings are $min\_npix = 27, min\_value = min\_delta = 2\sigma$. 
These mean that the clumps will include 27 voxels at least, the intensity of each voxel should be greater than $2\sigma$, and the minimum relative brightness of each clump should be greater than $2\sigma$. 
These values are selected to ensure that the resultant clumps are real and robust against noise while including as many faint clumps as possible to ensure reasonable statistics.
Due to the higher noise of the \thirteen line, we did not detect enough \thirteen clumps in the high latitude region for statistics.

To estimate the clumps' physical properties (velocity dispersion, size, excitation temperature, mass, surface density, and virial parameter), we use the following methods.

The column density and all mass related parameters are derived under a linear conversion from integrated intensity to column density, with $R_{31} = 0.31\pm0.11$ is the intensity ratio between CO (3-2) and CO (1-0) \citep{2024ARA&A..62..369S}, $X_\textrm{CO}=2.0 \times 10^{20} ~\textrm{cm}^{-2}~ (\textrm{K km s})^{-1}$ \citep{2013ARA&A..51..207B}, and $\mu_{\textrm{H}_2}=2.8$ \citep{2008A&A...487..993K}, which lead to a mass conversion factor of $\alpha_{\textrm{CO}32}=R_{31} ~\mu_{\textrm{H}_2} ~m_\textrm{H} ~X_\textrm{CO} = 1.38 ~M_{\odot} ~\textrm{pc}^{-2}~(\textrm{K km s})^{-1}$. 
The velocity dispersion and size are derived from the dendrogram result using a moment method \citep{2008ApJ...679.1338R}.
The excitation temperature is derived using the brightest voxel of \twelve  in the clump $T_{peak}$ and the radiation transfer with optical thick assumption as $$\frac{1}{exp(h\nu/k_B ~T_{ex}) -1 }= J(T_{ex}) \simeq k_B ~T_{peak}/h\nu + J(T_{bg})$$ and $$J(T_{bg}) = \frac{1}{exp(h\nu/k_BT_{bg}) -1 }  \ \ .$$
where the background is the cosmic microwave background as $T_{bg} = 2.726$ K, and the line frequency is $h\nu/k_B=16.596$ K.

\begin{figure}[tbp]
    \centering
    \includegraphics[width=\linewidth]{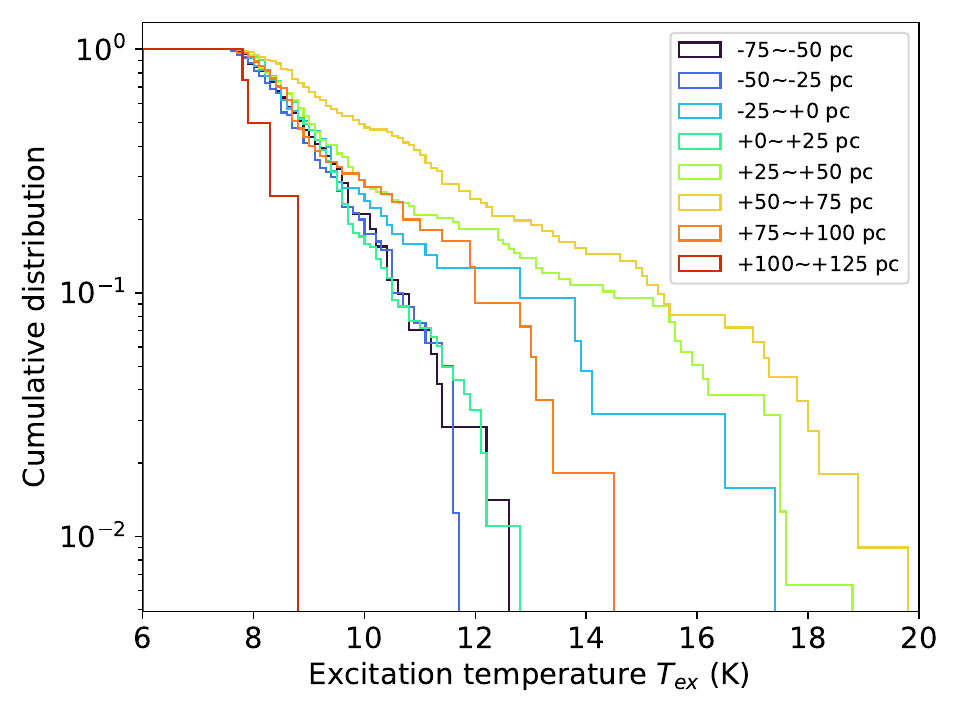}
    \caption{Cumulative distributions of the excitation temperature for clumps in the G309 region, grouped by their height $z$. The distributions are symmetric at about +50 pc. }
    \label{fig:multigroup}
\end{figure}

We analysed the distribution of these properties as functions of both the absolute height $|z|$ and the local height relative to the local centre of the spiral arm component $|z-z_0|$.
By comparing the distributions of these properties among different heights, we found that the distributions are more symmetric with the local $z=\bar z$ mid-plane rather than the global $z=0$ mid-plane. Fig. \ref{fig:multigroup} shows an example that the distributions are symmetric at about +50 pc for the G309 region.

We divided our clumps into low ($<50$ pc), median ($50\to100$ pc), and high ($100\to150$ pc) relative height groups to highlight their statistical difference, and named them as LRH, MRH, and HRH respectively. There are 1781, 707, and 30 clumps in these three groups for the G328 region, and 540, 230, and 39 clumps for the G309 region. The cumulative distributions of the physical parameters for these three groups of the two regions are shown in Fig. \ref{fig:parameters-G328} and \ref{fig:parameters-G309}, respectively. 

For each region, we compare the LRH and MRH distributions using one-sided two-sample Kolmogorov--Smirnov (K--S) tests and a two-sample Anderson--Darling (A--D) test; the resulting $p$-values are listed in Table \ref{tab:LRH-MRH-2arm}. The A--D test assesses the null hypothesis that the two samples are drawn from the same distribution, whereas each one-sided K--S test assesses a specified ordering of their cumulative distribution functions (CDFs). We interpret the listed $p$-values at the nominal significance level $p<0.05$, without accounting for multiple comparisons. Failure to reject a null hypothesis does not establish that the two distributions are identical. The A--D test does not determine the direction of a difference.

Excitation temperature and surface density show the clearest common trends: both the K--S and A--D tests indicate significant differences between LRH and MRH clumps in both regions, with the directional K--S results supporting higher LRH values. These results are consistent with MRH clumps being less likely to attain high excitation temperatures and surface densities, as shown in Fig. \ref{fig:parameters-G328} and \ref{fig:parameters-G309}. A similar tendency toward fewer high-mass clumps remains tentative, since the mass difference is significant only in the G328 region at the nominal 5\% level. Further analysis of the high-value tails with larger sample is needed to confirm this interpretation.

Velocity dispersion, radius, and virial parameter show no consistent directional trend across the two regions. More data would help determine whether this reflects regional differences or limited statistical sensitivity.

\begin{table}[tbp]
\centering
\caption{P-values of the K--S and A--D tests comparing MRH clumps to the LRH clumps.}
\label{tab:LRH-MRH-2arm}
\begin{tabular}{c|cc|cc}
\toprule
 & \multicolumn{2}{c|}{G328} & \multicolumn{2}{c}{G309}\\
Parameter & K--S & A--D & K--S & A--D\\
\midrule
$\sigma_v$      & 1.2e-2 (>) & 3.5e-3 & 4.2e-4 (<) & 7.0e-4\\
Radius          & 5.1e-2 (<) & 1.1e-1 & 9.8e-2 (>) & 1.6e-1\\
Temperature     & 2.5e-6 (<) & 1.0e-4 & 6.4e-3 (<) & 6.0e-4\\
Mass            & 3.8e-2 (<) & 2.0e-2 & 2.7e-1 (<) & 1.9e-1\\
Surface density & 1.1e-3 (<) & 7.0e-4 & 2.7e-4 (<) & 1.0e-4\\
Virial parameter& 6.3e-4 (>) & 1.0e-4 & 6.4e-2 (<) & 7.0e-4\\
\bottomrule
\end{tabular}
\tablefoot{For the one-sided K--S test, a small $p$-value for $(<)$ supports smaller MRH values than LRH values, whereas a small $p$-value for $(>)$ supports larger MRH values.}
\end{table}

\section{Discussion}
\label{sect:Discussion}
\subsection{Integrated vertical properties of molecular disk}
\label{sect:Discussion:intergrate}
We have fitted the \twelve vertical distributions with two components (wide and narrow). Here we compare this result  with other reports from the literature: the old CfA-1.2m result \citep{1994ApJ...436L.173D} and the new PMODLH-13.7m MWISP result \citep{2021ApJ...910..131S}.

Both the CfA and MWISP results are derived from $^{12}\mathrm{CO}\ J=1\to0$ observations conducted in the first Galactic quadrant. 
The CfA result is toward three regions ($l=30\degr,40\degr,50\degr$) and the MWISP result is an average among a large portion of tangent regions. 
Although the CfA survey encompasses a broader range in Galactic latitude ($|b|<10$\degr), when the vertical extent of the Galactic thick disk is taken into account, the MWISP data set provides results that are generally of superior quality with respect to those obtained from the CfA observations.
Both results consistently show that the molecular disk observed by $^{12}\mathrm{CO}\ J=1\to0$ can be separated into two components with FWHMs of approximately 85 pc and 280 pc, which refer to the thin and thick molecular disk.
The MWISP results also show that the FWHM of thin disk will drop from about 90 pc at $l=52\degr$ to about 60 pc at $l=16\degr$ for the tangent region.
However, these tangent region in the first quadrant is close to the Galactic bar.
Their results are likely not only represent the inner Galactic disk but also include the effect of the Galactic bar.

We have observed two spiral-arm tangent regions in the fourth Galactic quadrant. This allows our line-of-sight to follow the spiral arms rather than go through the inter-arm tangent region like the CfA result. Hence, we include and average among much more material while keeping the same physical resolution across our mapping region.
The FWHM of the two components in our results are about 25 pc and 100 pc, similar in the two tangent regions and much smaller than the CfA and MWISP result.
We also note that, the two components of G328 region share similar center about +65 pc but not for the G309 region, which may reflect that the narrow components are more dominated by the local effect.

We propose two possible explanations for our two components as follows: the first explanation related them as two sub-component of the thin disk (85 pc), in which our narrow one (25 pc) traces the spiral arm, and the wide one (100 pc) traces the thin disk.

The second explanation is the two components we measure are still the thin and thick disk but just thinner. 
Because our observations employ the higher rotational transition $J=3\to2$, whereas the CfA and MWISP surveys utilize the lower $J=1\to0$ transition, our data are preferentially sensitive to, and therefore predominantly trace, the higher-density components of the molecular gas. 
This is similar to the different thicknesses of the atomic gas disk and the molecular gas disk. 
Also the HI absorption disk is narrower than the HI emission disk at all Galactic radii \citep{2023ARA&A..61...19M}.
In other words, as the characteristic density of the observed medium gradually increases (from warm neutral medium and cold neutral medium to those traced by $^{12}\mathrm{CO}\ J=1\to0$ and $J=3\to2$), the inferred thickness gradually decreases, but the thickness ratio of the two components still remains at about 1:3 to 1:4. 
At even higher densities, the ATLASGAL inner-Galactic dust clumps exhibit an exponential scale height of 25 pc, corresponding to a FWHM of 60 pc \citep{2018MNRAS.473.1059U}.
\citet{2024ApJ...963L...9X} reported a paucity of dense dust cores in high-latitude Planck cold clumps in the solar neighbourhood, at distances of 100–400 pc and absolute $z$-heights of 80–200 pc.
In combination with the results presented in Sect. \ref{sect:result:clump}, this suggests that, at large heights above the Galactic midplane, the density distribution of the interstellar medium drops faster at the high-density end.
Rigorous confirmation or refutation of this implication will require more detailed studies and higher-resolution observations, preferably in the solar neighbourhood to ensure sufficient physical resolution.

\subsection{Statistical properties of clumps}
\label{sect:Discussion:clumps}
Our results indicate that the environments located above and below the Galactic disk are associated with variations in clump physical properties, with clear trends observed  in excitation temperature and surface density.
At the same time, compared with the lower disk at inner Galaxy, the outer Galaxy also contains less matter and shows similar environmental changes with the high-height region.
Taking into account this similarity, we compared our results between low- and median-height clumps with those derived between the inner and outer Galaxy.
Comparing the $J=2\to1$ surveys of inner and outer Galaxy (SEDIGISM and OGHReS), it is reported that there is no significant difference in surface densities between inner- and outer-Galaxy clumps, with a K-S test $p$-value of 0.52 \citep{2024MNRAS.528.4746U}.
Conversely, our result shows that there is a statistically significant difference in surface densities between low and median relative height clumps, as shown in Fig. \ref{fig:parameters-G328}(e) and \ref{fig:parameters-G309}(e).
We note that this discrepancy may be induced by an observational bias since the clumps in both SEDIGISM and OGHReS not only have different Galactocentric distances but also different distances to the Sun. 
This means that their clumps have different physical resolutions, which instead is uniform in our sample.

\subsection{Impact on the study of the Galactic gradients}
Traditionally, studies of Galactic gradients (e.g. isotopologues ratio, gas to dust ratio etc. against Galactocentric distance) have selected dense cores from Galactic plane surveys for follow-up observations, using targeted single-pointing data to derive relationships between parameters of interest and Galactocentric distance. 
This approach implicitly assumes that variations arising from the scale height above the Galactic plane are negligible compared to those driven by Galactocentric distance. 
However, our findings demonstrate that this assumption breaks down. 
Although Galactocentric distances are numerically much larger than vertical heights, the average height of these dense gas samples is actually comparable to the thickness of the dense disk gas layer. 
Consequently, variations induced by vertical offset from the mid-plane become non-negligible relative to those along the Galactocentric radial gradient.
This may explain part of the large uncertainties in the studies of Galactic gradients.

\section{Summary}
We have mapped two spiral arm tangent regions (G328 and G309), covering up to $b=[-2\degr, +2\degr]$ (or $|z|\sim200$ pc near the tangent points). The noise level of the \twelve line is about 0.4 K [$T_{MB}$] with a velocity channel of 0.5 \kms.
Here we summarize our main result as follows:
\begin{enumerate}
\item The vertical distributions around the spiral arm tangent regions are fitted with two Gaussian components, a narrow one with an FWHM thickness of 25 pc, and a wide one with a FWHM of about 100 pc.
Two explanations are proposed: they are two sub-components of the thin disk, or they trace the thin and thick disks which show a smaller thickness when observed with higher transition of CO rotational lines. The latter indicates that at large heights above the Galactic mid-plane, the density distribution of the interstellar medium exhibits a steeper gradient.
\item Clumps in the two tangent regions are searched and divided into groups according to their relative height.
Compared with the low-relative-height clumps, the median-relative-height clumps have lower excitation temperature and surface density, with fewer high values in the distribution. A similar tentative trend is found for the mass.
\end{enumerate}

\begin{acknowledgements}
This work is partly supported by the Collaborative Research Center 1601 (SFB 1601 sub-project B1) funded by the Deutsche Forschungsgemeinschaft (DFG, German Research Foundation) – 500700252. 
We acknowledge the anonymous referee for suggestions which significant improved our paper.
The data was collected under the Atacama Pathfinder EXperiment (APEX) Project, led by the Max Planck Institute for Radio Astronomy at the ESO La Silla Paranal Observatory.
This research made use of \textit{Astropy} (a community-developed core \textit{Python} package for Astronomy \citep{2013A&A...558A..33A, 2018AJ....156..123A}), \textit{NumPy} \citep{2020Natur.585..357H}, \textit{SciPy} \citep{2020NaMet..17..261V}, and \textit{Matplotlib} \citep{2007CSE.....9...90H}.
\end{acknowledgements}

\bibliographystyle{aa}
\bibliography{export-bibtex}

\newpage

\begin{appendix} 
\section{Pixel efficiency model of the LASMA receiver}
\label{app:eff_model}
As shown in \citet[Sect. 2.2.2]{2022PhDT.........4M}, the 7 pixels of the LASMA receiver have different gains depending on individual pixels and the receiver's derotation angle (recorded as dewar angle $\theta_{dewar}$).
Here, we derived the pixel efficiency model to precisely convert our data from the antenna temperature scale \Ta to the main-beam brightness temperature scale \Tmb.

The vast majority of LASMA data is toward the bright \twelve and \thirteen lines of several giant molecular clouds \citep[e.g.][and more in prep.]{2021A&A...650A.164M, 2024A&A...692A.114N}.
These data are obtained using the classical method that all 7 pixels cover the same sky area 7 times.
We use all these data observed from 2022 to 2024 to derive our efficiency model.

\begin{figure}[htbp!]
    \subfloat[]{
        \begin{minipage}{\linewidth}
        \centering
        \includegraphics[width=\textwidth]{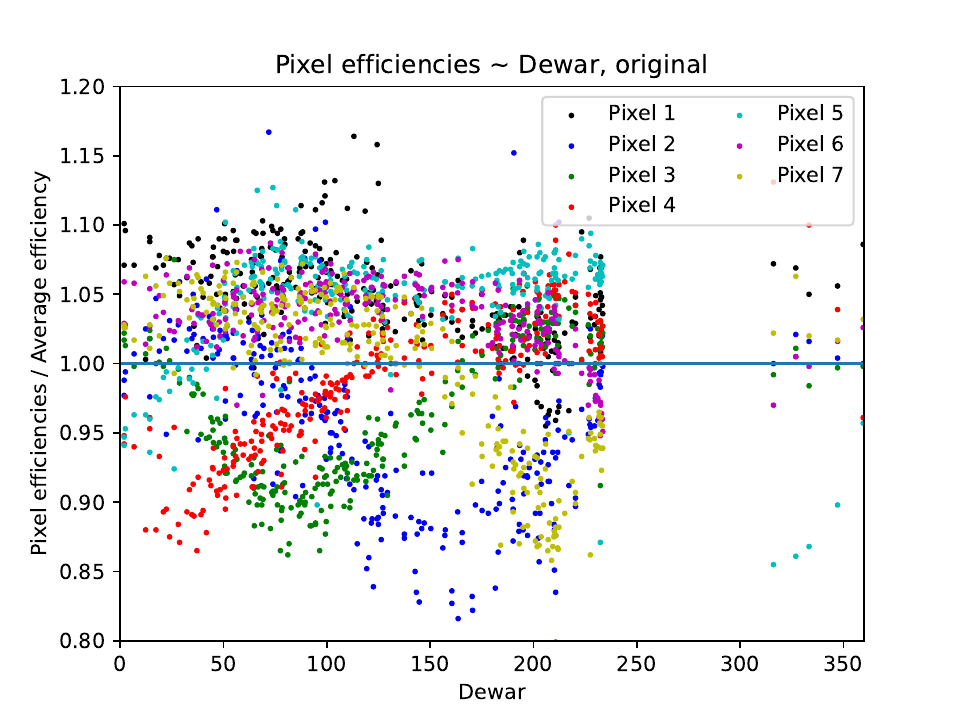}
    \end{minipage} } 
    
    \subfloat[]{
        \begin{minipage}{\linewidth}
        \centering
        \includegraphics[width=\textwidth]{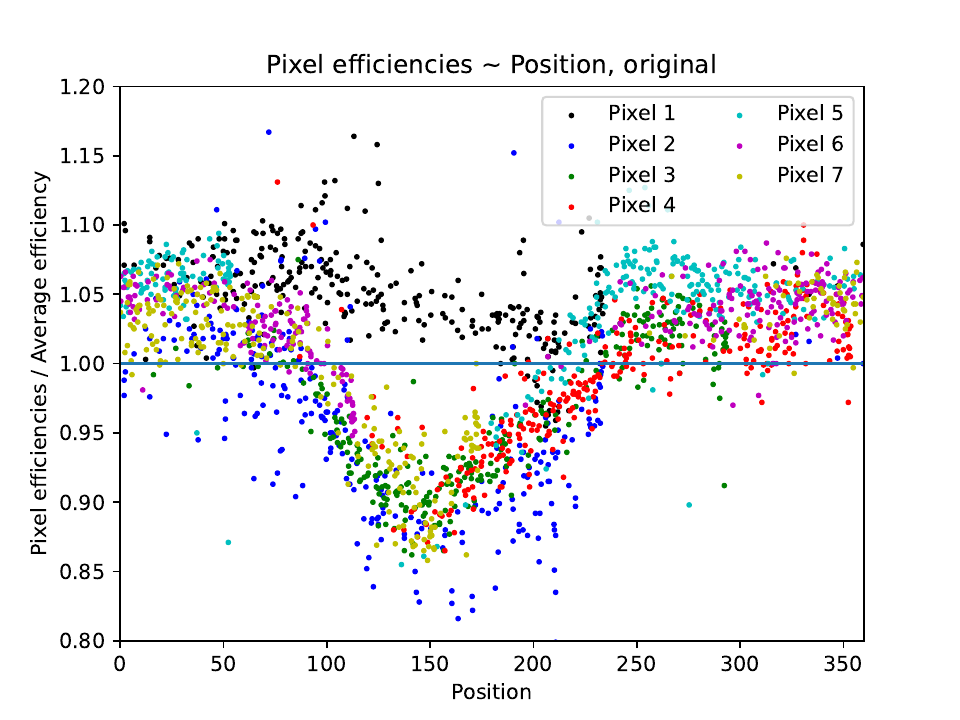}
    \end{minipage} } 
    
    \subfloat[]{
        \begin{minipage}{\linewidth}
        \centering
        \includegraphics[width=\textwidth]{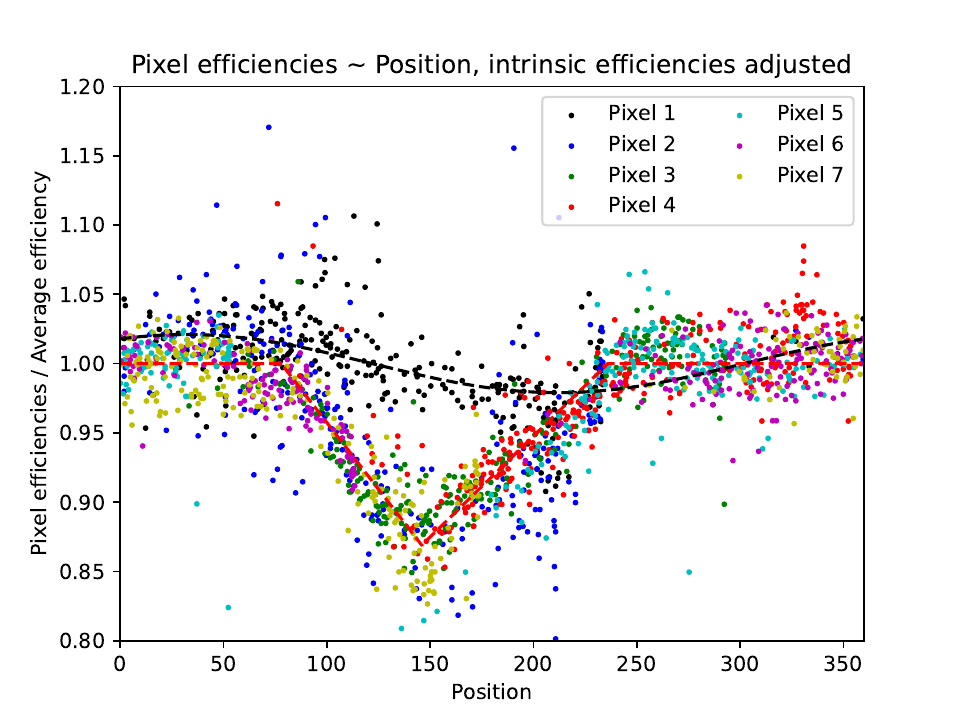}
    \end{minipage} } 
    \caption{The relationship between (a) origin efficiencies and dewar angle, (b) origin efficiencies and position angle, and (c) pixel-corrected efficiencies and position angle. Different colours represent different pixels.}
    \label{fig:lasma:efficiency}
\end{figure}

\begin{figure}[htbp!]
    \subfloat[]{
        \begin{minipage}{\linewidth}
        \centering
        \includegraphics[width=0.9\linewidth]{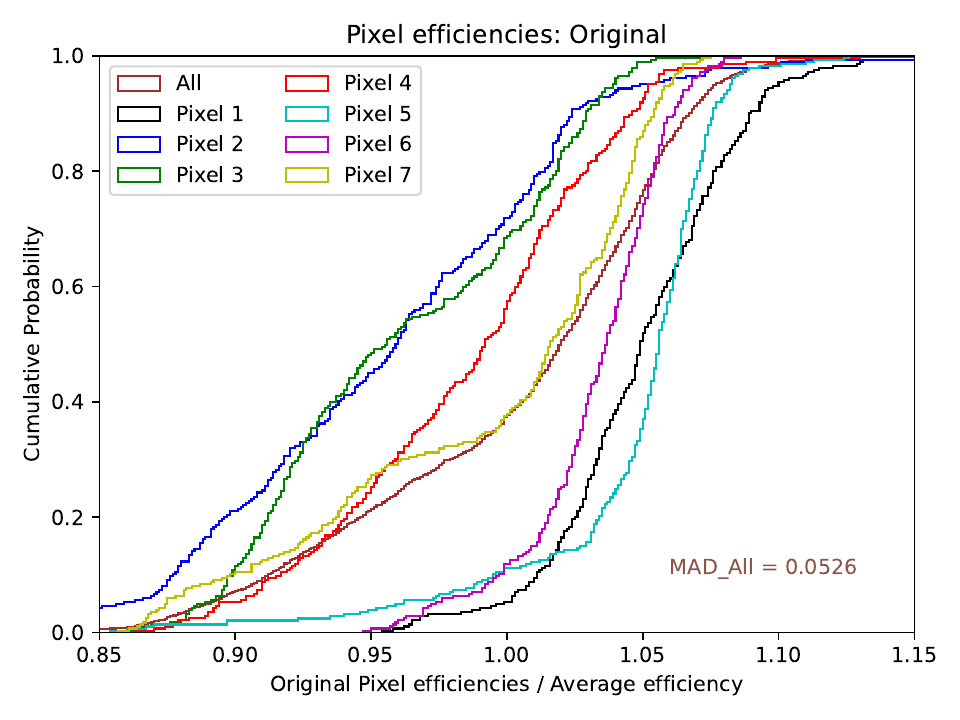}
    \end{minipage} } 
    \hfill
    \subfloat[]{
        \begin{minipage}{\linewidth}
        \centering
        \includegraphics[width=0.9\linewidth]{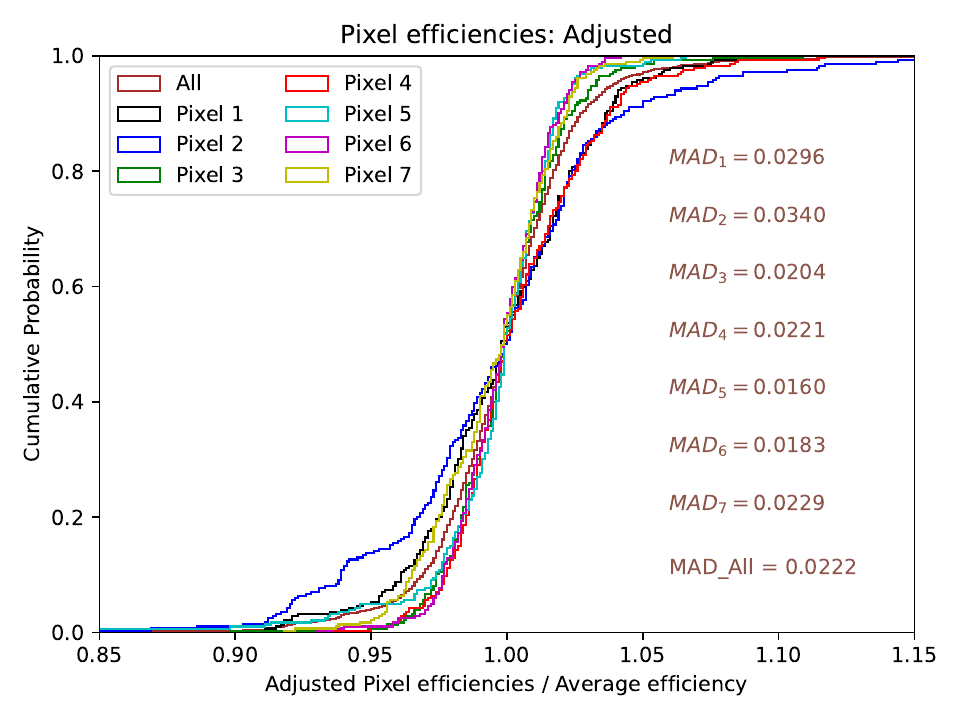}
    \end{minipage} } 
    \caption{The cumulative distribution of (a) origin efficiencies, and (b) residual efficiencies after applying the model for each pixel. Their corresponding standard deviation computed from the median absolute deviation (MAD) are shown.}
    \label{fig:lasma:CDF}
\end{figure}

We note the fact that corresponding voxels representing the same sky area and velocity channel and observed by different pixels in different receiver derotation angles should have the same main beam brightness temperature \Tmb.
For each scan, we derive the "average" data cube using all pixels and weighted by noise, the same as what one would do for normal surveys. This is regarded as the true value from the sky. Then we also derive the data cube for each pixel. By comparing the cube from one pixel with the "average" cube, we get the intensity ratio for every pixel of one scan (one dewar angle).
The median intensity ratio over all bright voxels in the cube for each scan is used as the pixel's relative efficiency in that scan.
By repeating these for all the aforementioned LASMA scans from 2022 to 2024, we get the relative efficiency of 286 bright scans for all 7 receivers.
The relation between these relative efficiencies and the dewar angle $\theta_{dewar}$ is shown in Fig. \ref{fig:lasma:efficiency}(a), while Fig. \ref{fig:lasma:CDF}(a) shows the cumulative distribution of these origin efficiencies.
The origin efficiencies have a relative deviation of 5.3\% (computed from the median absolute deviation, MAD) and systematic differences among pixels.

By converting from the dewar angle $\theta_{dewar}$ to the position angle $\theta_{pos} = \theta_{dewar} + (i-2)\times 60^\circ$ for outer pixels ($i$ is the pixel number from 2 to 7), it is clear that the centre pixel has a sinus dependence (black dots) while the outer 6 pixels are affected by a triangle-like "absorption" as Fig. \ref{fig:lasma:efficiency}(b).
They are fitted by 7 intrinsic pixel efficiencies multiplied by a sinus or the absorption, with the intrinsic-corrected efficiencies shown as Fig. \ref{fig:lasma:efficiency}(c).

The fitted model values are as follows:
The relative efficiency of the centre pixel is 
\begin{equation}
\eta_1' = 1.052 + 0.021 \sin{(\theta_{dewar}+58\degr)}
\label{equ:1}
\end{equation}
and the relative efficiency of the outer pixel $i$ (2 to 7) is
\begin{equation}
\eta_i' = \begin{cases}
        \eta_i ~~~~~~~~~~~~~~, \text{ if } 0\degr \leq \theta_{pos} < 78\degr ~ \text{ or } ~ 236\degr \leq \theta_{pos} < 360\degr ; \\
        \eta_i \times(1-0.131\frac{\theta_{pos}-78\degr}{68\degr}) ~~~~~~~, \text{ if } 78\degr \le \theta_{pos} < 146\degr ;\\
        \eta_i \times(1-0.131\frac{236\degr-\theta_{pos}}{90\degr}) ~~~~, \text{ if } 146\degr \leq \theta_{pos} < 236\degr.
\end{cases}
\label{equ:2}
\end{equation}
where the intrinsic efficiencies $\eta_i$ for pixel $i$ are $\eta_2=0.997, \eta_3=1.015, \eta_4=1.014, \eta_5=1.057, \eta_6=1.043, \eta_7=1.038$, respectively.

The cumulative distribution of the model residual is shown in Fig. \ref{fig:lasma:CDF}(b). 
Now there is no systematic bias among the 7 pixels. 
The MAD-derived standard deviation is reduced to 2.2\%.

The aforementioned efficiencies convert the data from the \Ta scale into a homogeneous calibrated scale.
To further convert to the main beam brightness temperature \Tmb scale, we considered that the LASMA main beam efficiency measurements toward planets are calibrated using the centre pixel with a dewar angle range of $50\degr \sim 100\degr$ \citep[Sect. 2.2.2, Page 52]{2022PhDT.........4M}.
Using an average dewar angle of 75\degr, the relative efficiency for the $\eta_{MB}$ calibration is about $\eta_{cal}=1.067$.

Data of pixel $i$ in the main beam brightness temperature \Tmb scale is finally derived by
\begin{equation}
T_{MB,i} = \frac{\eta_{fw}} {\eta_{MB}} \frac{\eta_{cal}} {\eta_i'}  T_{A,i}^* =  \frac{0.95 \times 1.067} {0.77} \frac{T_{A,i}^*} {\eta_i'} 
\end{equation}
where $\eta_{fw}=0.95$ is the forward efficiency of the LASMA receiver since 2018, $\eta_i'$ is the relative efficiency derived from Eq. \ref{equ:1} or \ref{equ:2}, and $\eta_{MB}$ is the main beam efficiency which is monthly tested and reported by the APEX operation team\footnote{\url{https://www.apex-telescope.org/telescope/efficiency/}}.
Data in this paper were observed in 2023, hence we accept $\eta_{MB} \simeq 0.77\pm0.06$.

\section{The RMS noise maps of the \twelve line}
\begin{figure}[htbp!]
    \centering
    \subfloat[]{
        \includegraphics[height=0.38\textheight]{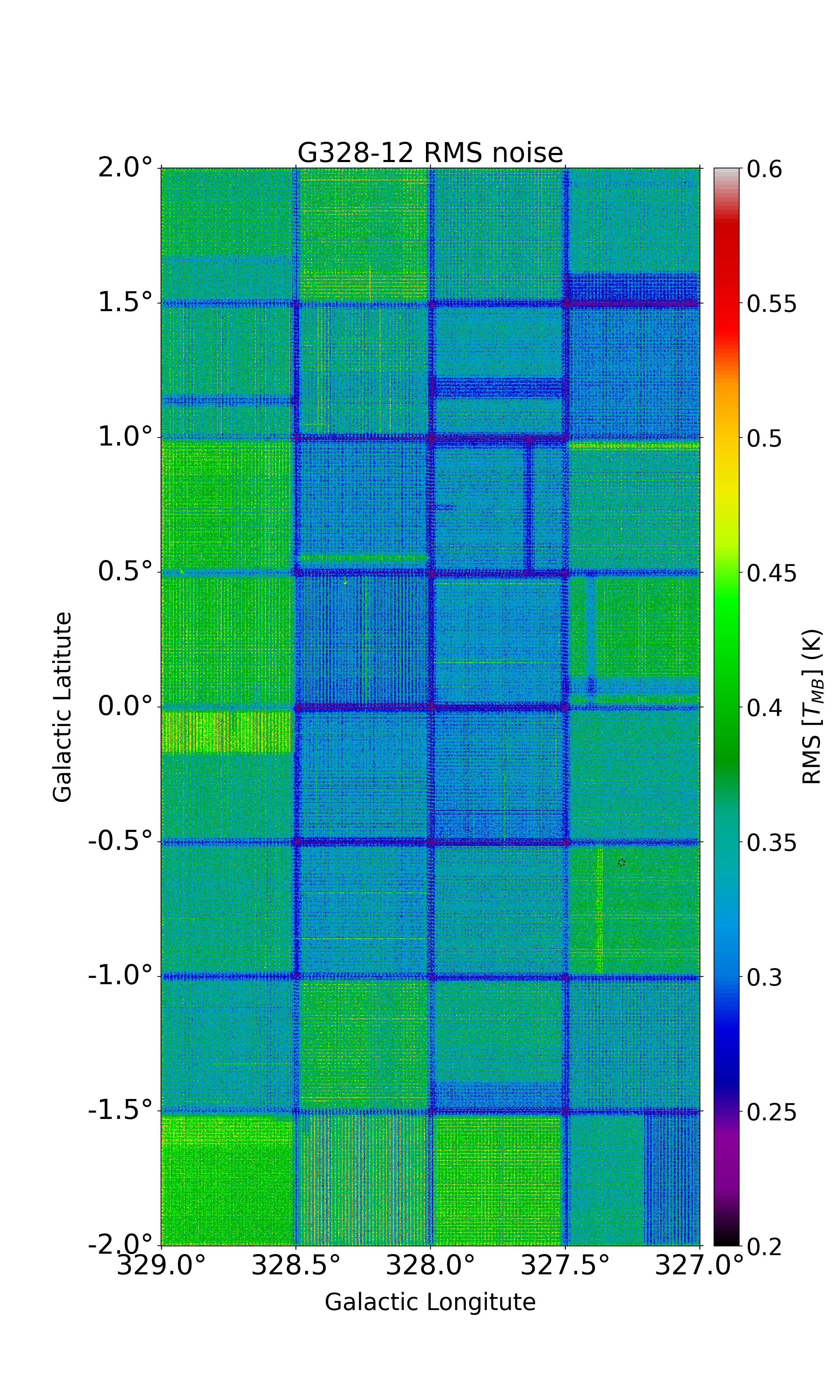}
        } 
        
    \subfloat[]{
        \includegraphics[height=0.38\textheight]{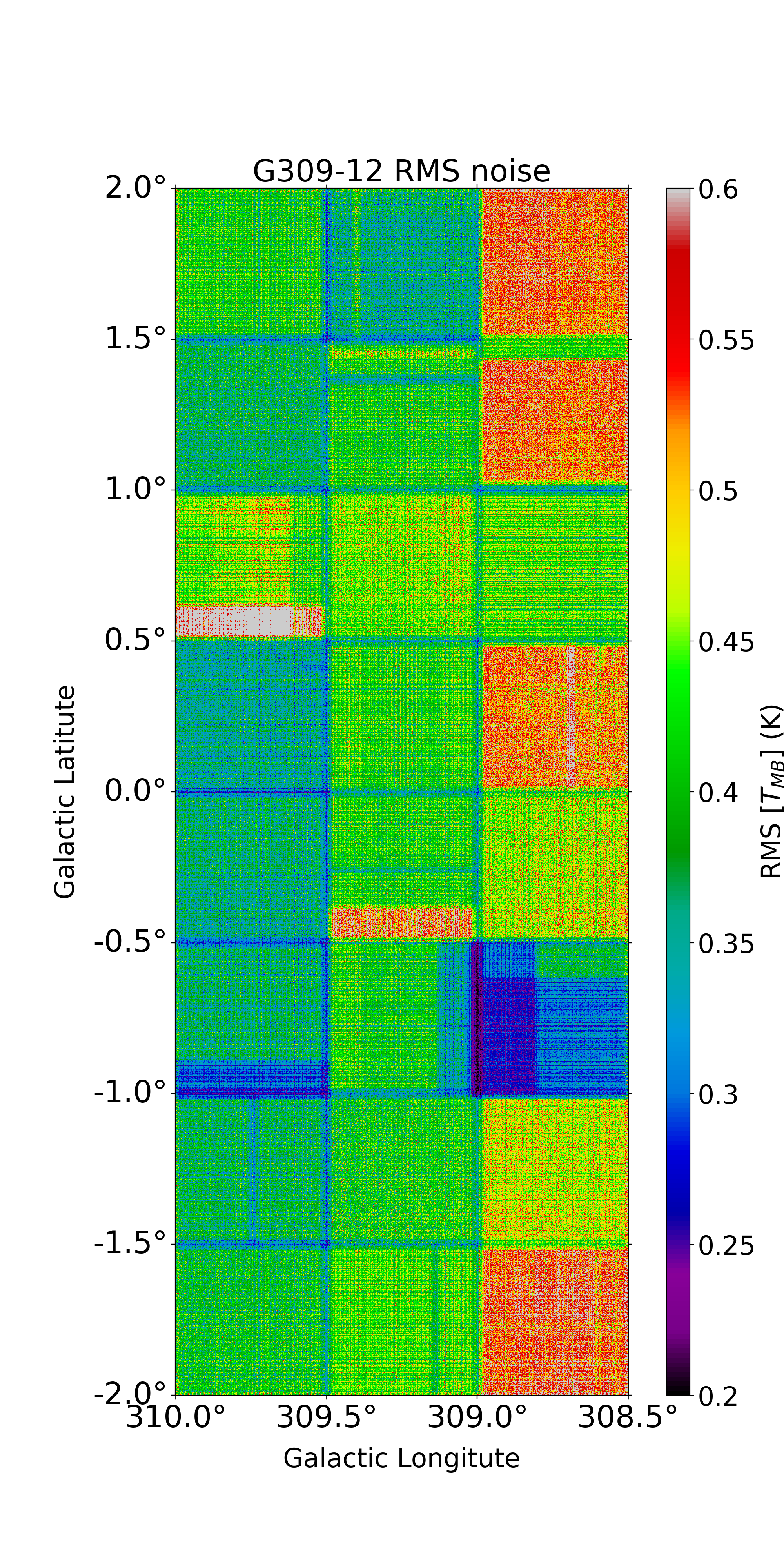}
        } 
    \caption{The RMS noise maps of the \twelve line for the two regions. Note the different colour scales of these two maps. The overlaps between fields led to lower noise in the grid parts.}
    \label{fig:rms-maps}
\end{figure}

The RMS noise maps of the \twelve line for the two regions are shown in \Tmb scale as Fig. \ref{fig:rms-maps}.

\FloatBarrier
\newpage
\onecolumn
\section{Integrated intensity maps of spiral arms}
\label{app:m0map}
\begin{figure*}[h]
    \centering

    \subfloat[]{
        \includegraphics[width=0.3\linewidth]{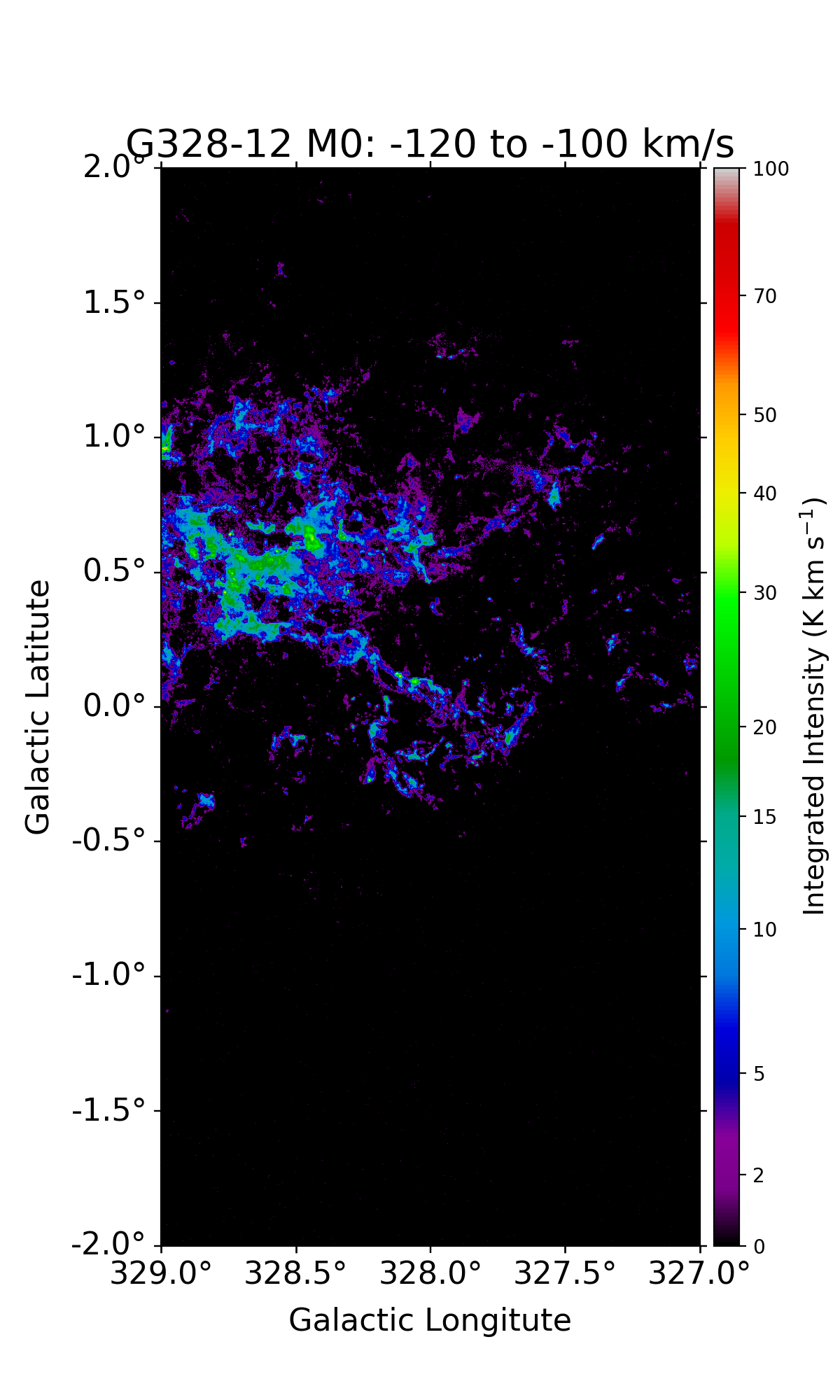}
    }
    \subfloat[]{
        \includegraphics[width=0.3\linewidth]{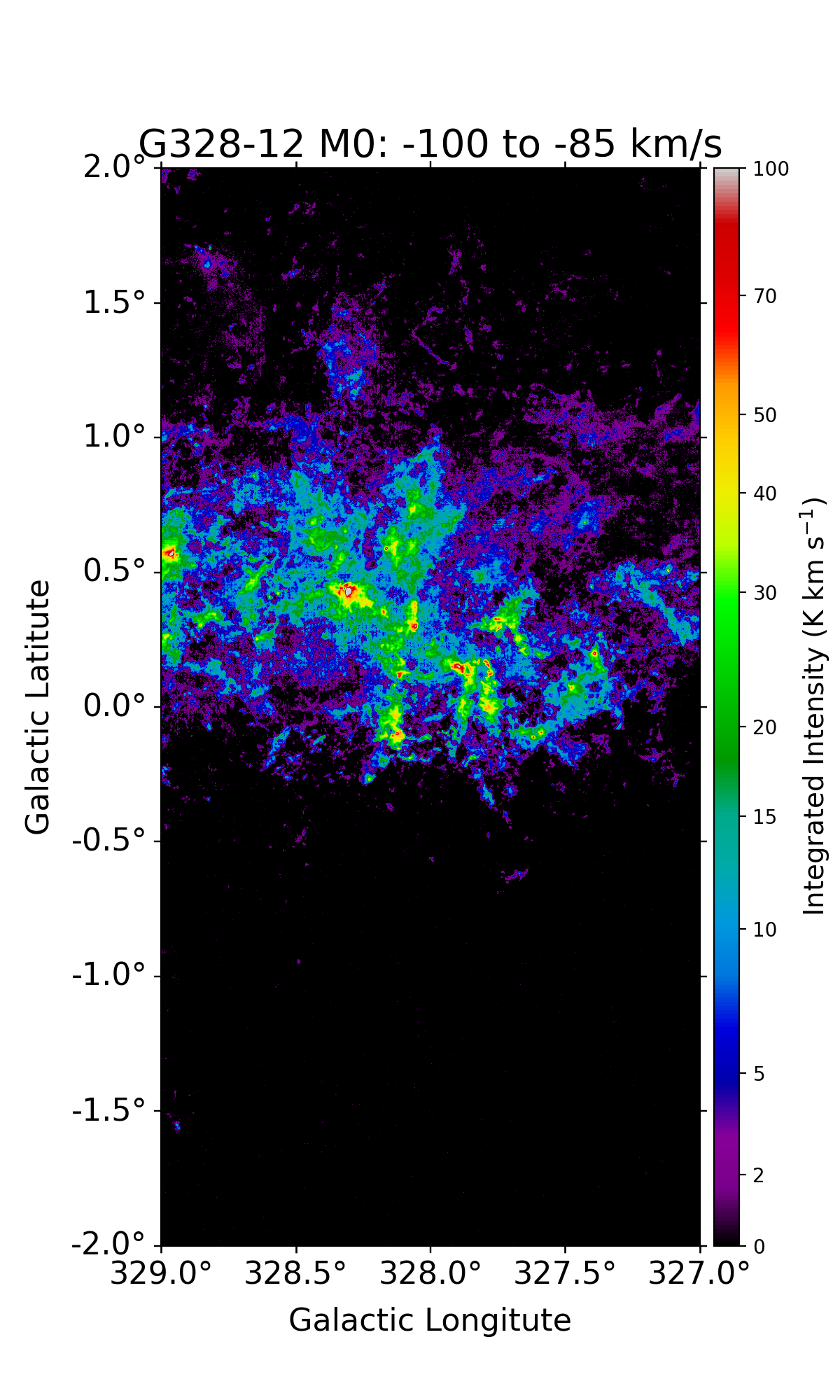}
    }
    \subfloat[]{
        \includegraphics[width=0.3\linewidth]{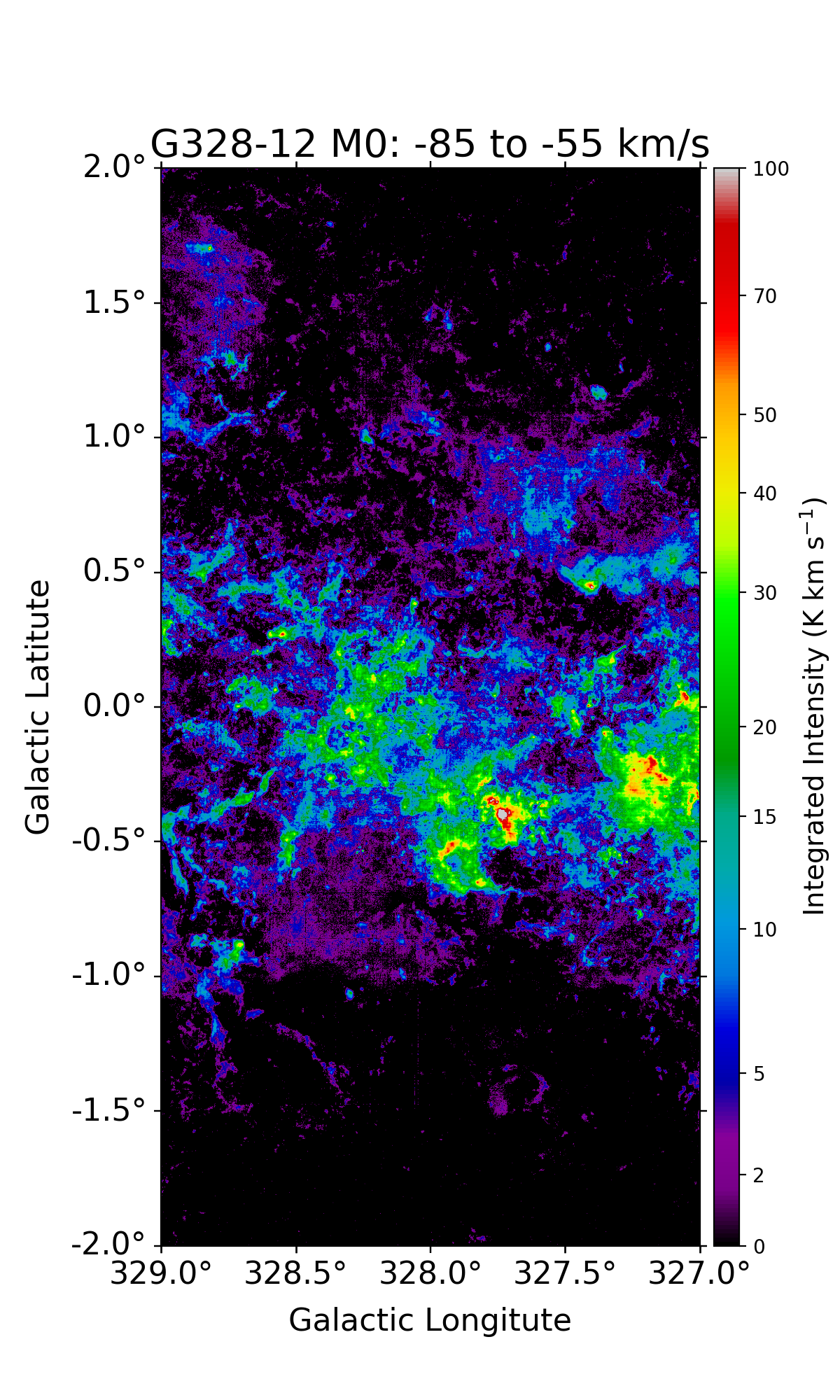}
    }
    
    \subfloat[]{
        \includegraphics[width=0.24\linewidth]{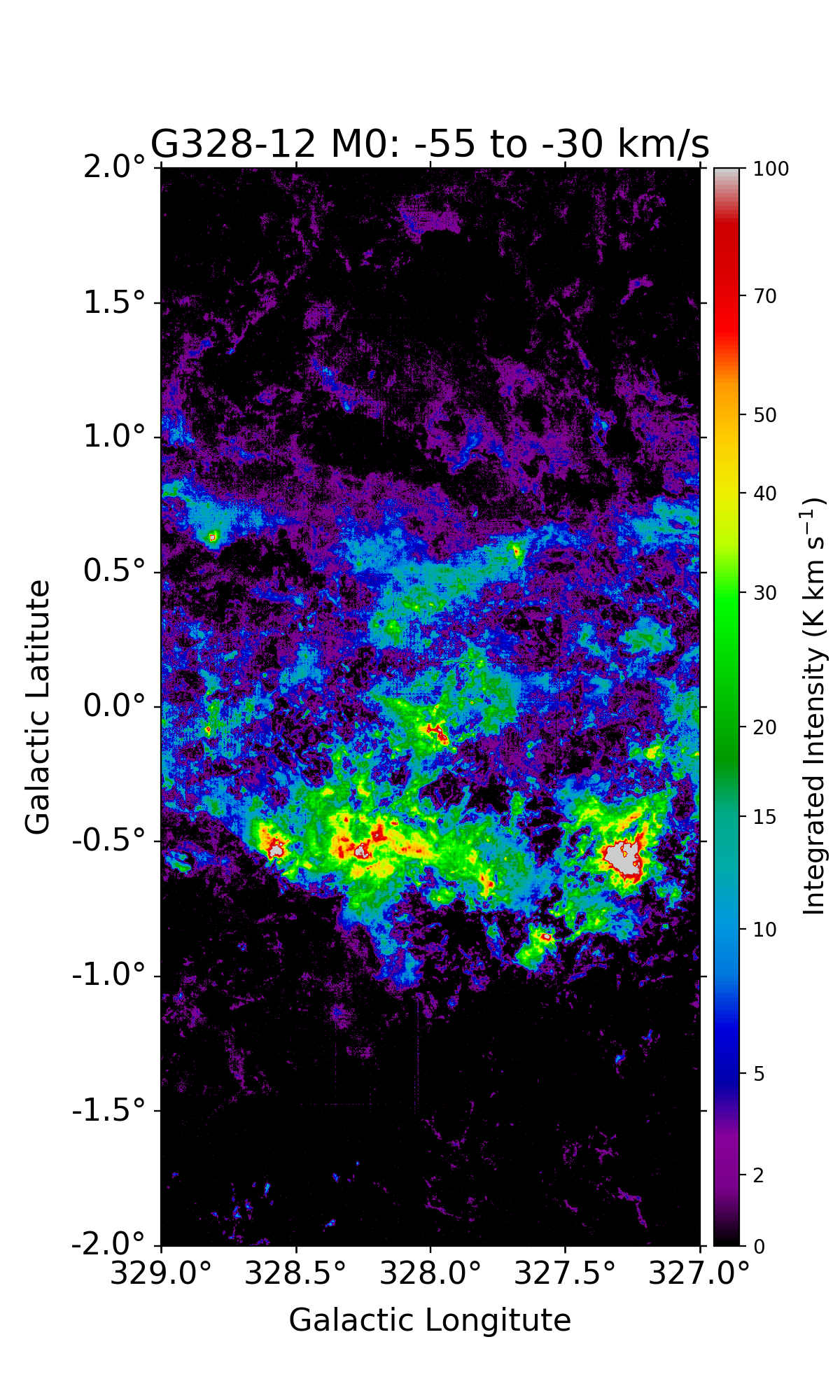}
    }
    \subfloat[]{
        \includegraphics[width=0.24\linewidth]{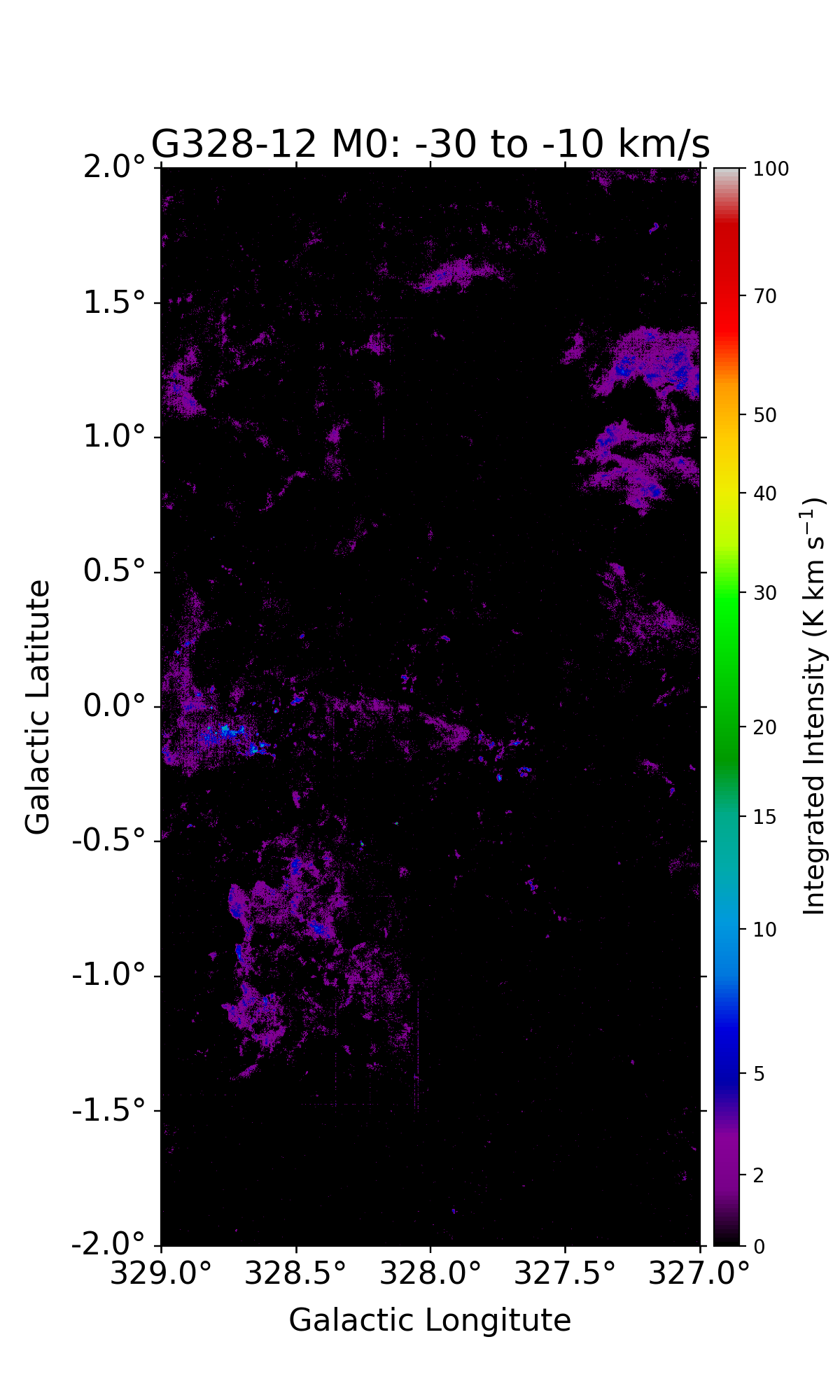}
    }
    \subfloat[]{
        \includegraphics[width=0.24\linewidth]{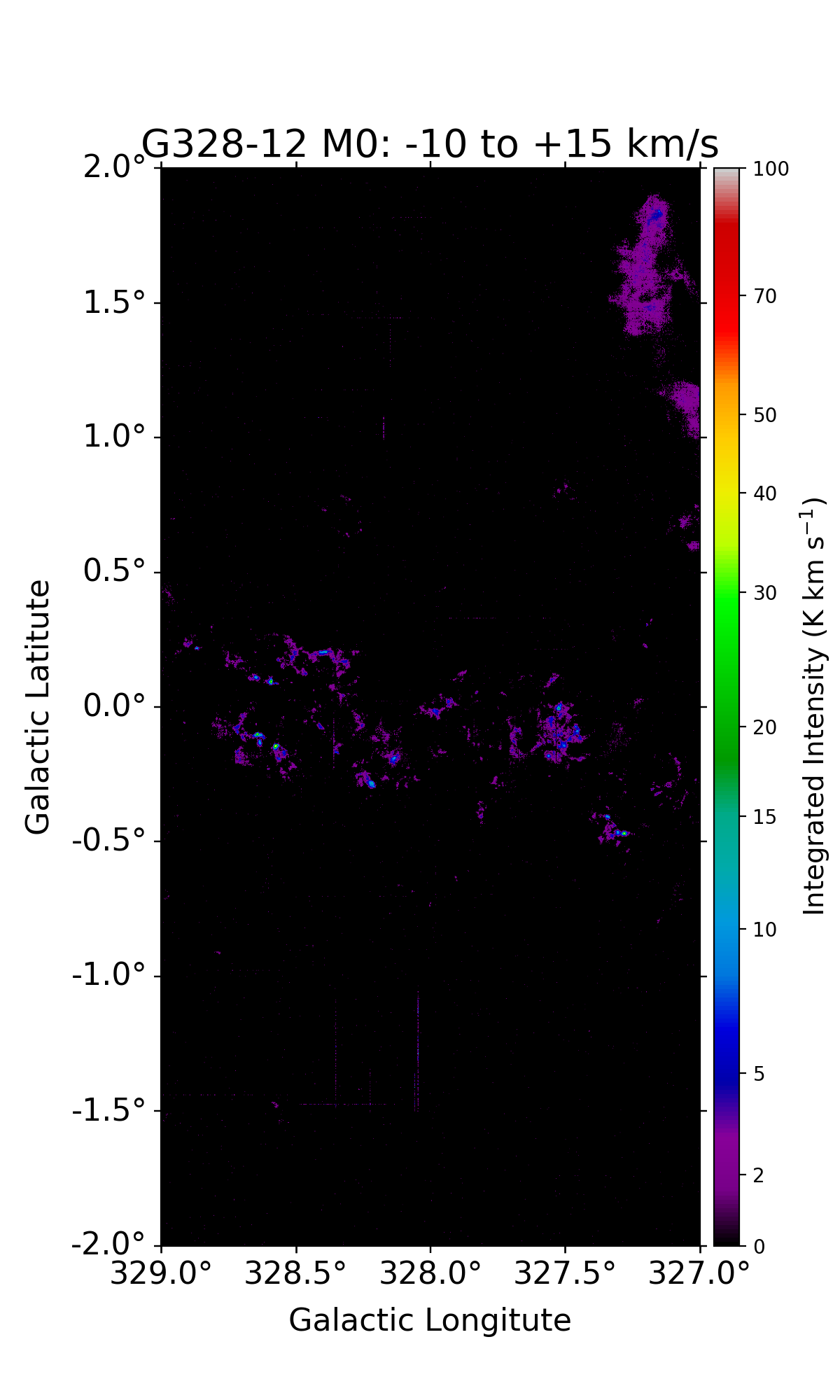}
    }
    \subfloat[]{
        \includegraphics[width=0.24\linewidth]{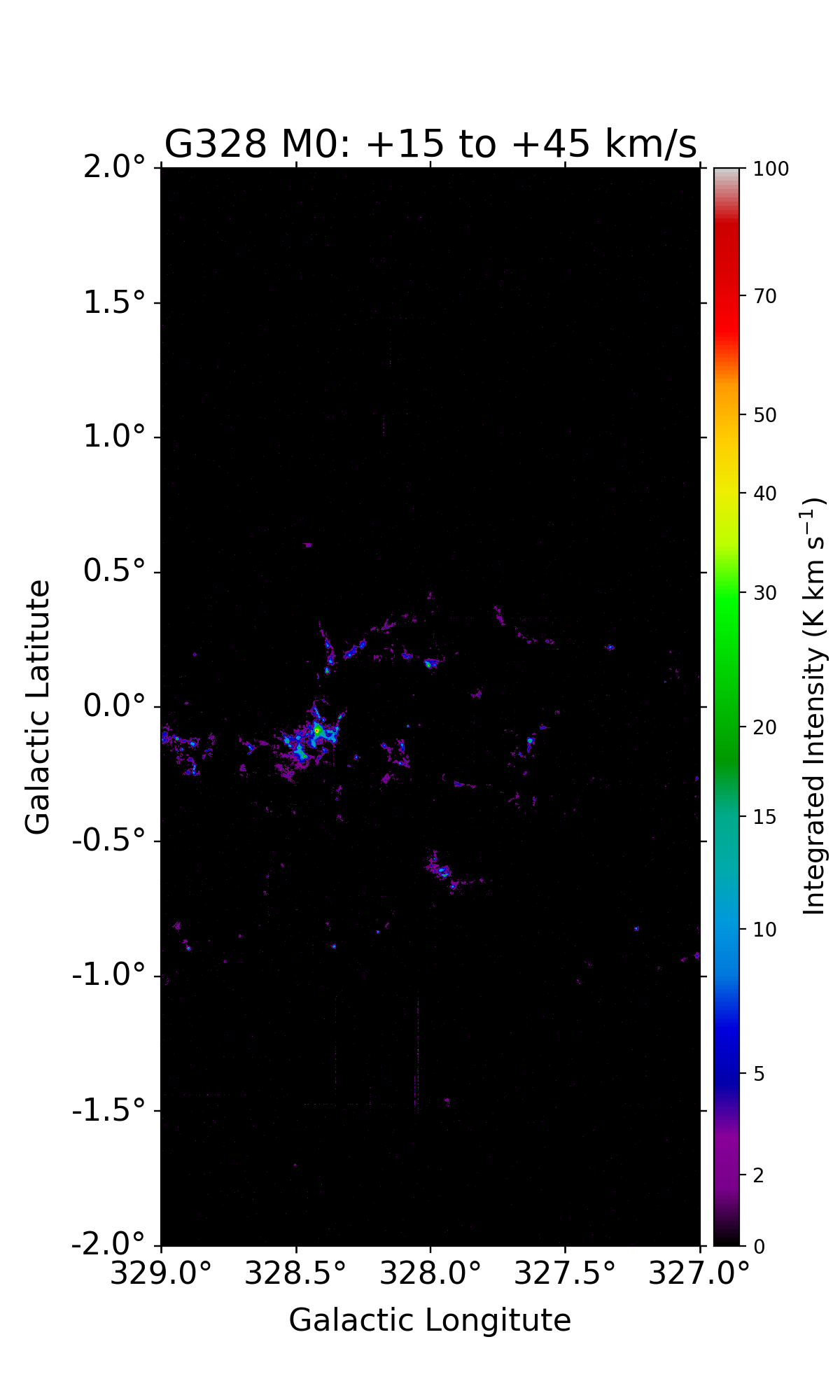}
    }

    \caption{The moment-zero maps of the \twelve line
    for different spiral arms of the G328 region.}
    \label{fig:m0maps-G328}
\end{figure*}

\begin{figure*}[h]
    \subfloat[]{
        \includegraphics[width=0.3\linewidth]{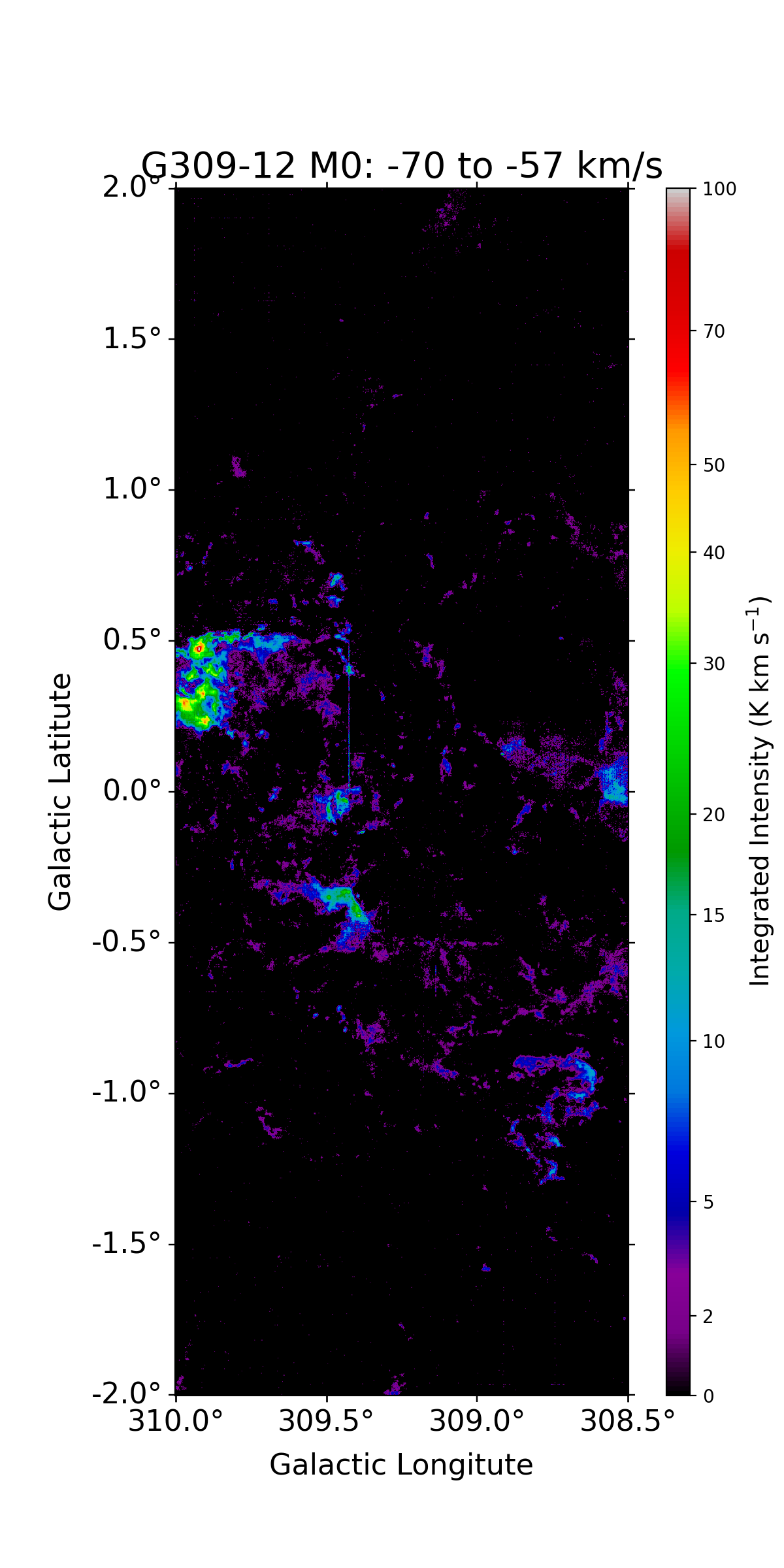}
    }
    \subfloat[]{
        \includegraphics[width=0.3\linewidth]{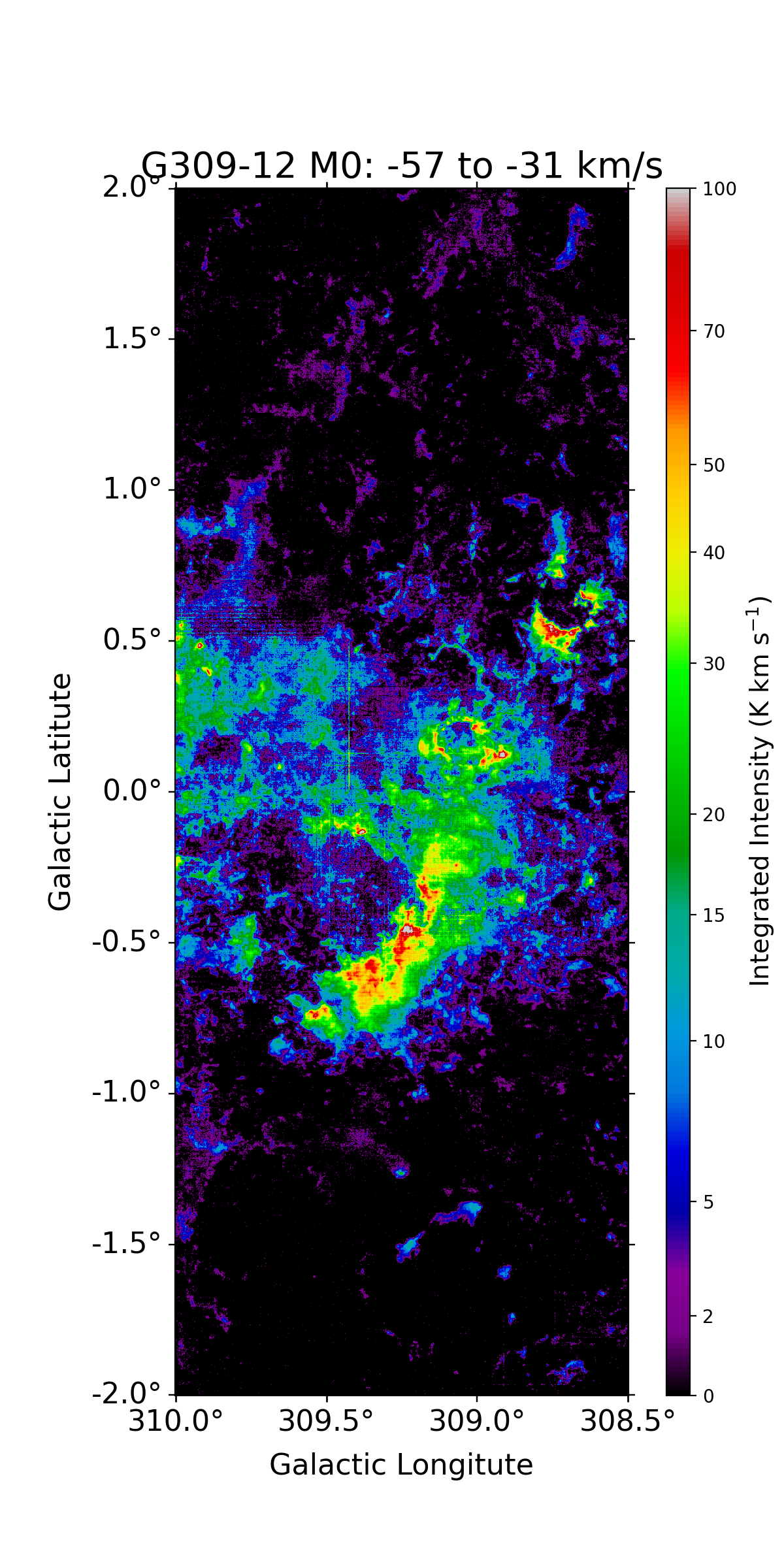}
    }
    \subfloat[]{
        \includegraphics[width=0.3\linewidth]{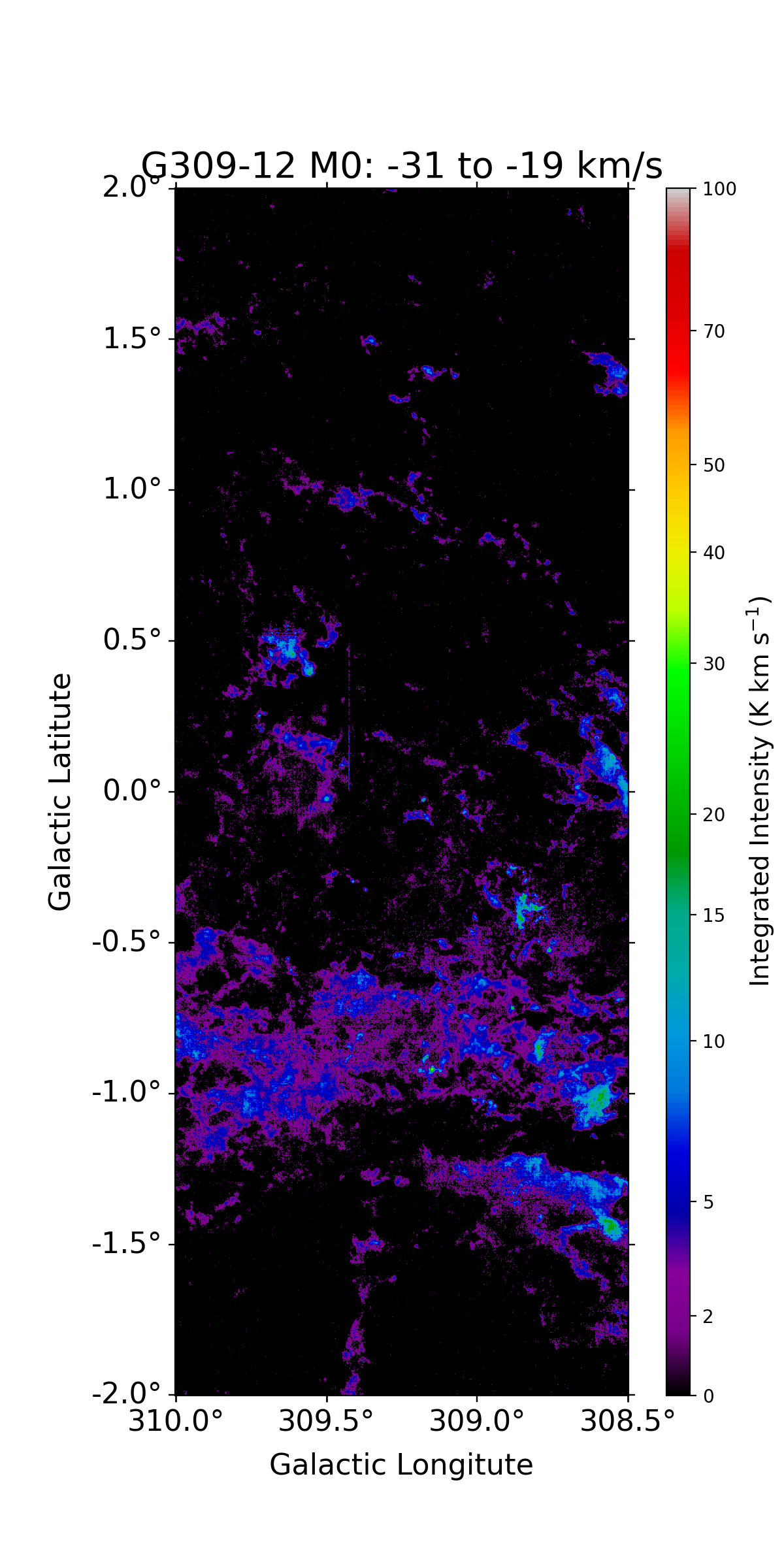}
    }
    
    \subfloat[]{
        \includegraphics[width=0.3\linewidth]{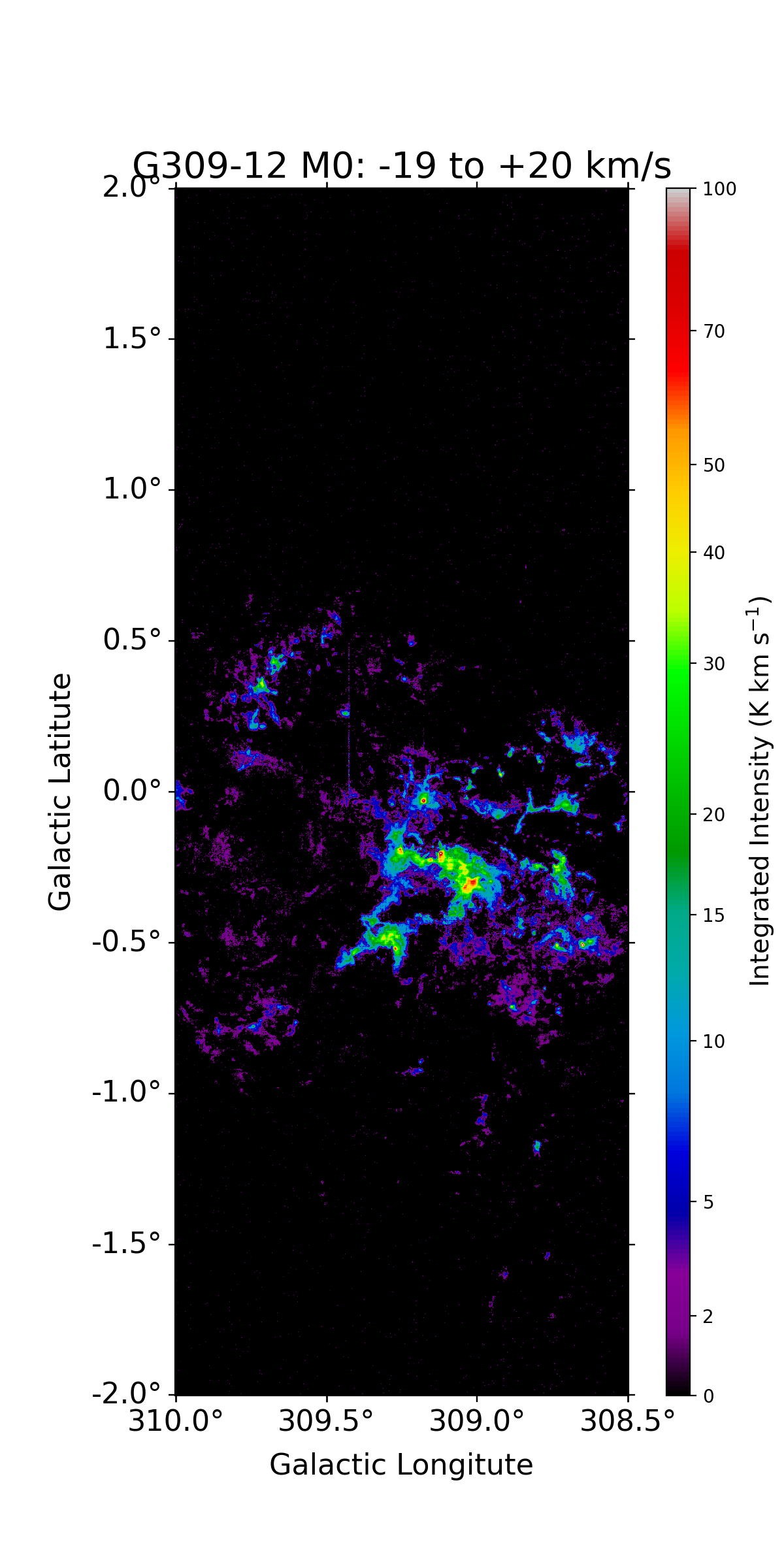}
    }
    \subfloat[]{
        \includegraphics[width=0.3\linewidth]{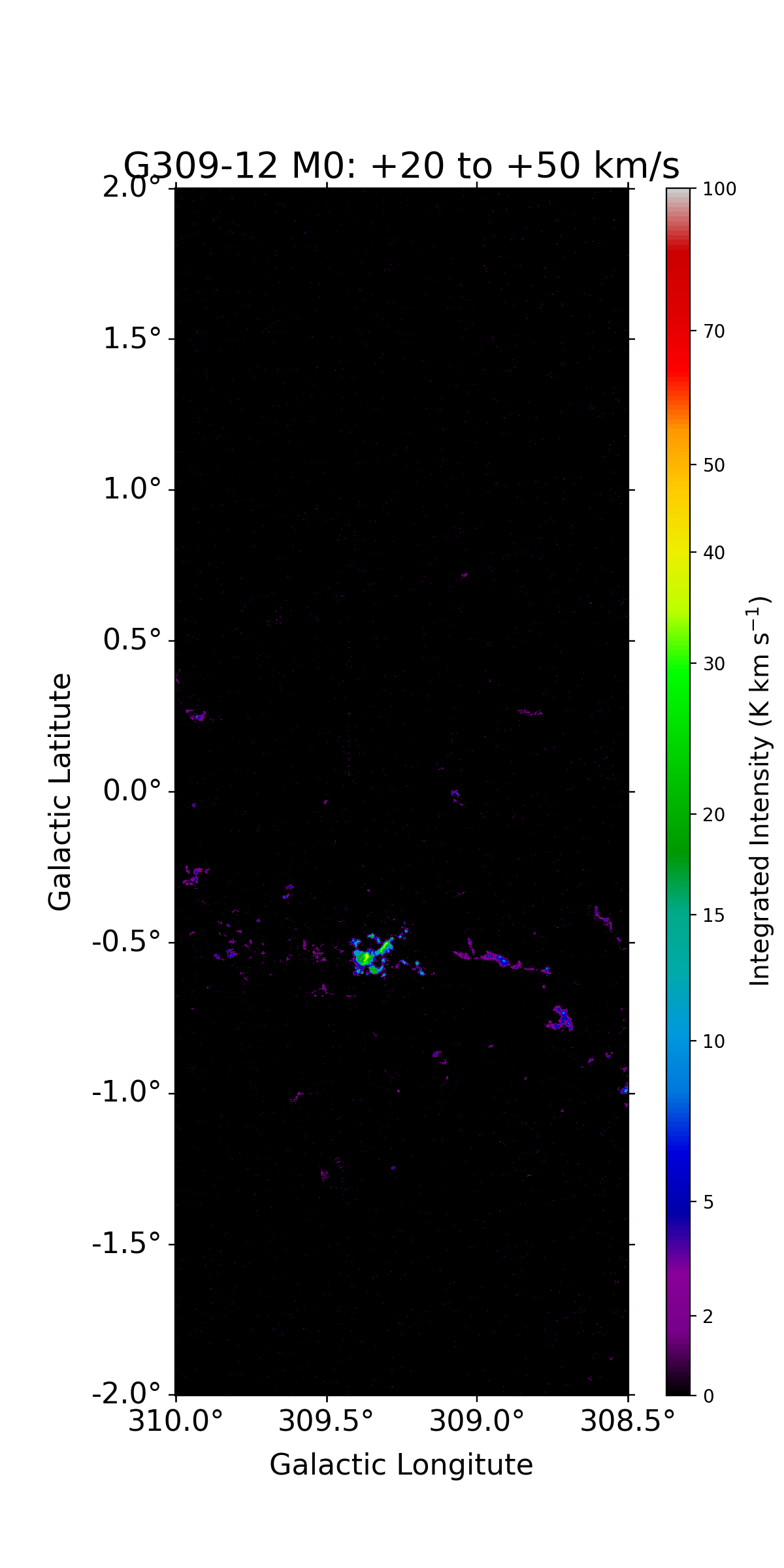}
    }

\caption{Same as Fig. \ref{fig:m0maps-G328} but for the G309 region.}
\label{fig:m0maps-G309}
\end{figure*}

The \twelve integrated intensity maps of different spiral arms for the G328 and G309 regions are shown as Fig. \ref{fig:m0maps-G328} and \ref{fig:m0maps-G309}, respectively.

\FloatBarrier
\newpage
\section{Cumulative distributions of clumps' physical parameters}
\begin{figure}[h]
    \centering
    \begin{minipage}{0.48\linewidth}
        \centering
        \subfloat[]{\includegraphics[width=\textwidth]{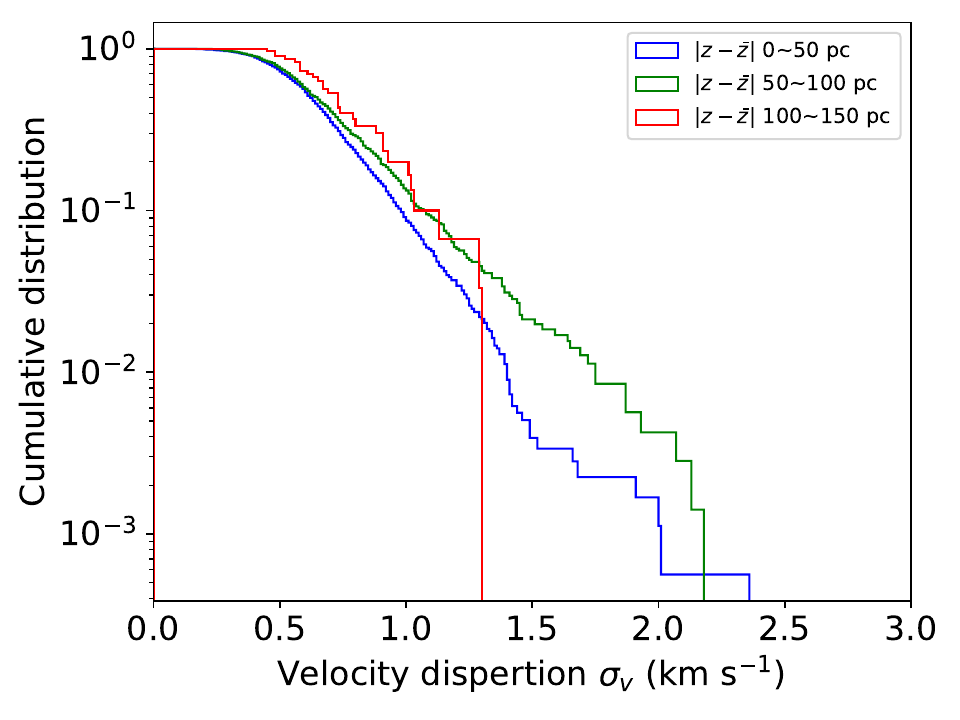}}
    \end{minipage}
    \hfill
    \begin{minipage}{0.48\linewidth}
        \centering
        \subfloat[]{\includegraphics[width=\textwidth]{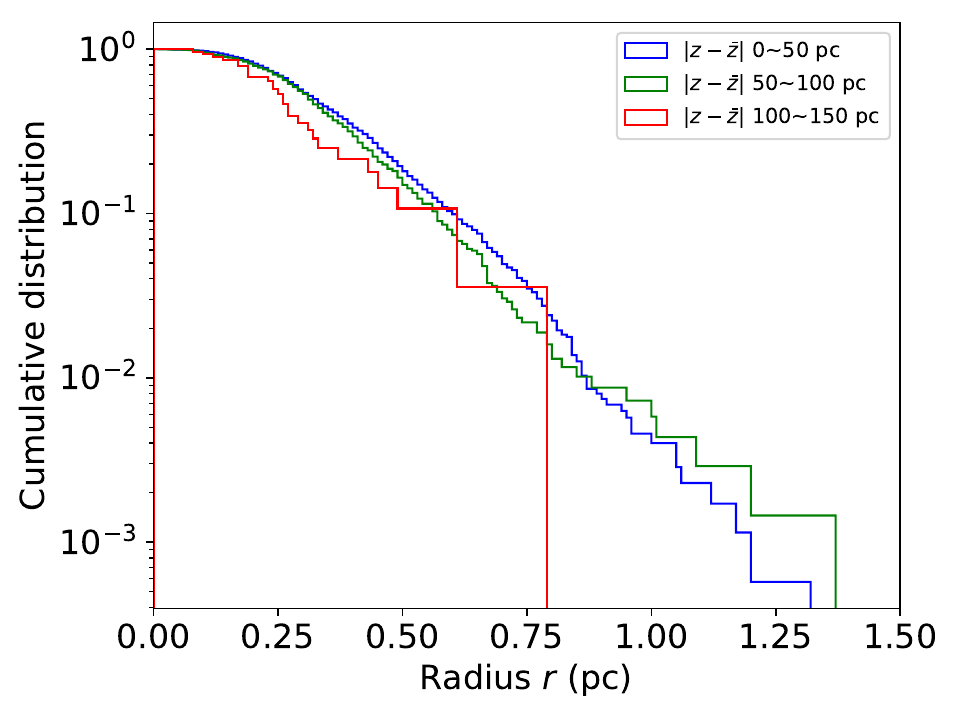}}
    \end{minipage}
    
    \begin{minipage}{0.48\linewidth}
        \centering
        \subfloat[]{\includegraphics[width=\textwidth]{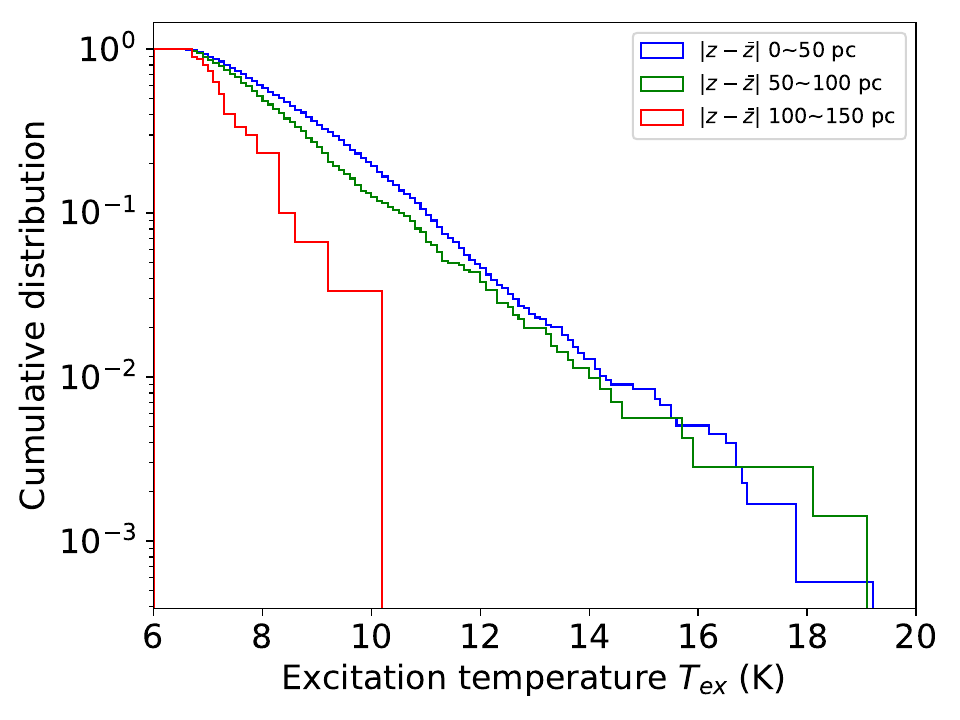}}
    \end{minipage}
    \hfill
    \begin{minipage}{0.48\linewidth}
        \centering
        \subfloat[]{\includegraphics[width=\textwidth]{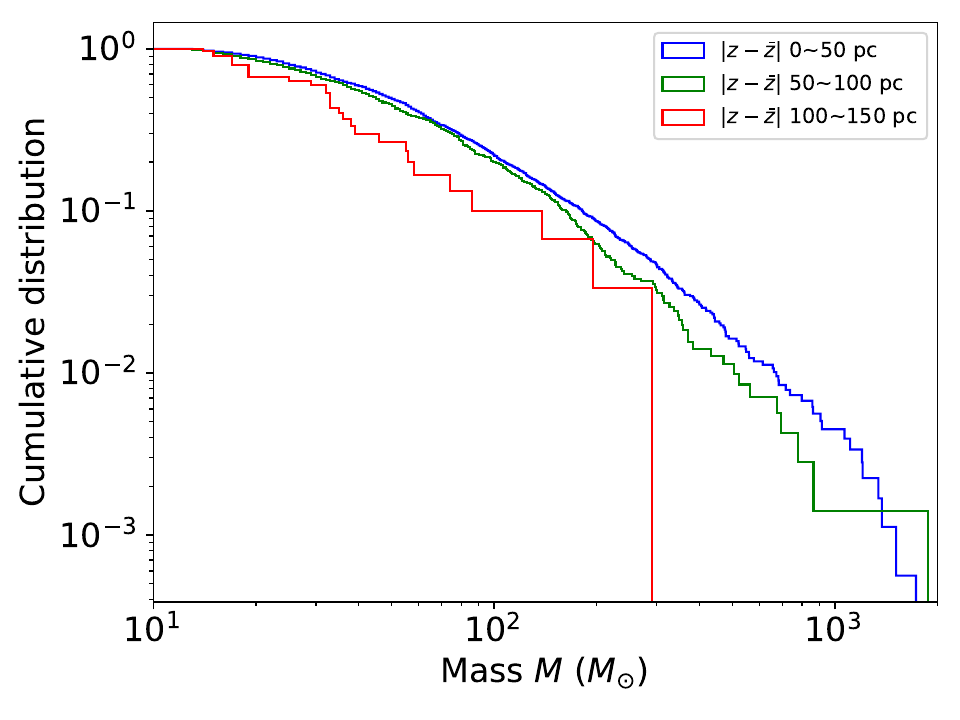}}
    \end{minipage}
        
    \begin{minipage}{0.48\linewidth}
        \centering
        \subfloat[]{\includegraphics[width=\textwidth]{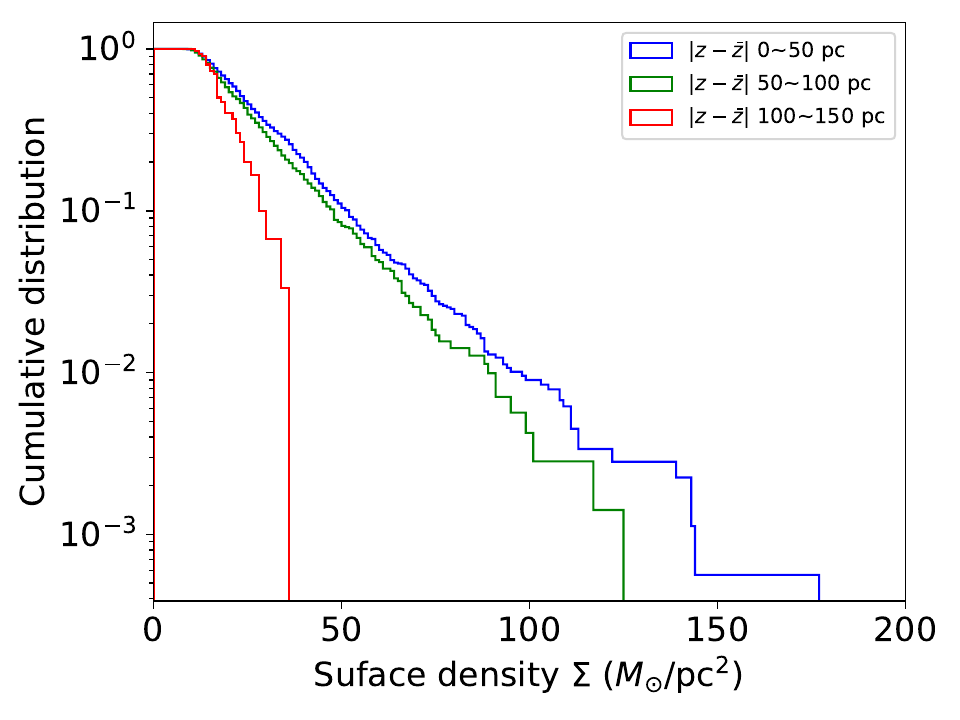}}
    \end{minipage}
    \hfill
    \begin{minipage}{0.48\linewidth}
        \centering
        \subfloat[]{\includegraphics[width=\textwidth]{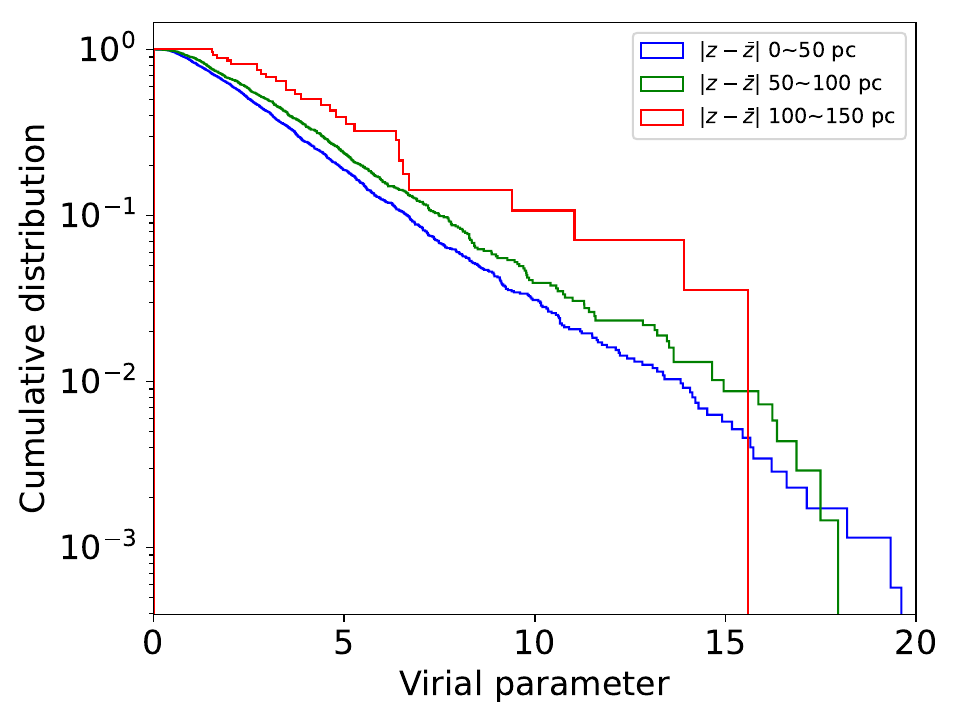}}
    \end{minipage}
    
    \caption{Cumulative distributions of velocity dispersion, $1\sigma$ radius, excitation temperature, mass, surface density, and virial parameter for clumps in the G328 region, divided into three groups by local relative height $|z-\bar z|$.}
    \label{fig:parameters-G328}
\end{figure}

\begin{figure*}[p]
    \centering
    \begin{minipage}{0.48\linewidth}
        \centering
        \subfloat[]{\includegraphics[width=\textwidth]{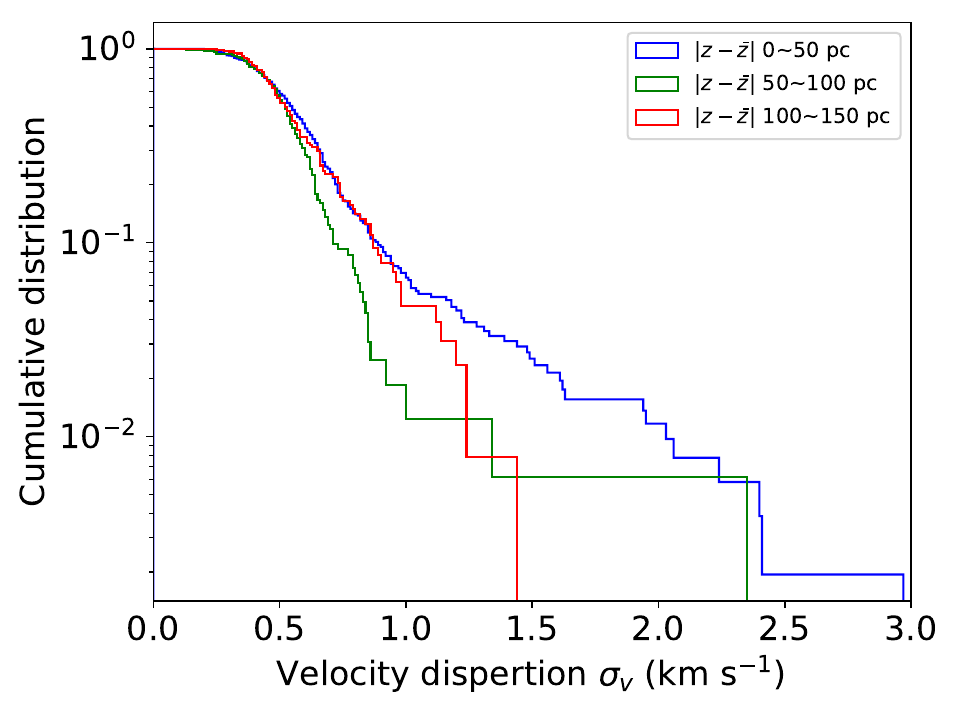}}
    \end{minipage}
    \hfill
    \begin{minipage}{0.48\linewidth}
        \centering
        \subfloat[]{\includegraphics[width=\textwidth]{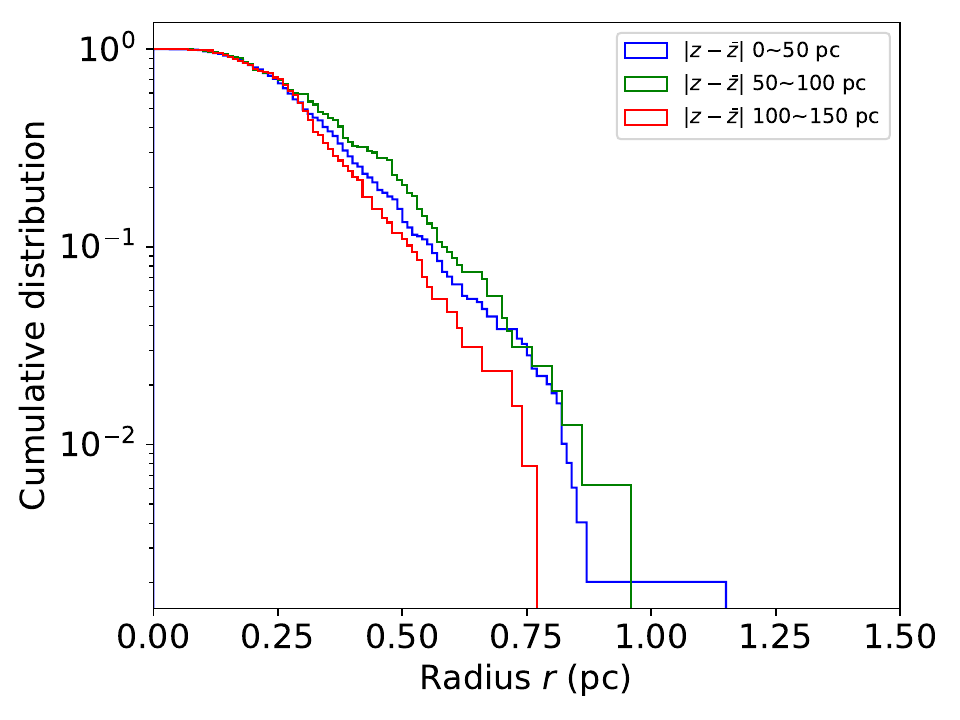}}
    \end{minipage}
    
    \begin{minipage}{0.48\linewidth}
        \centering
        \subfloat[]{\includegraphics[width=\textwidth]{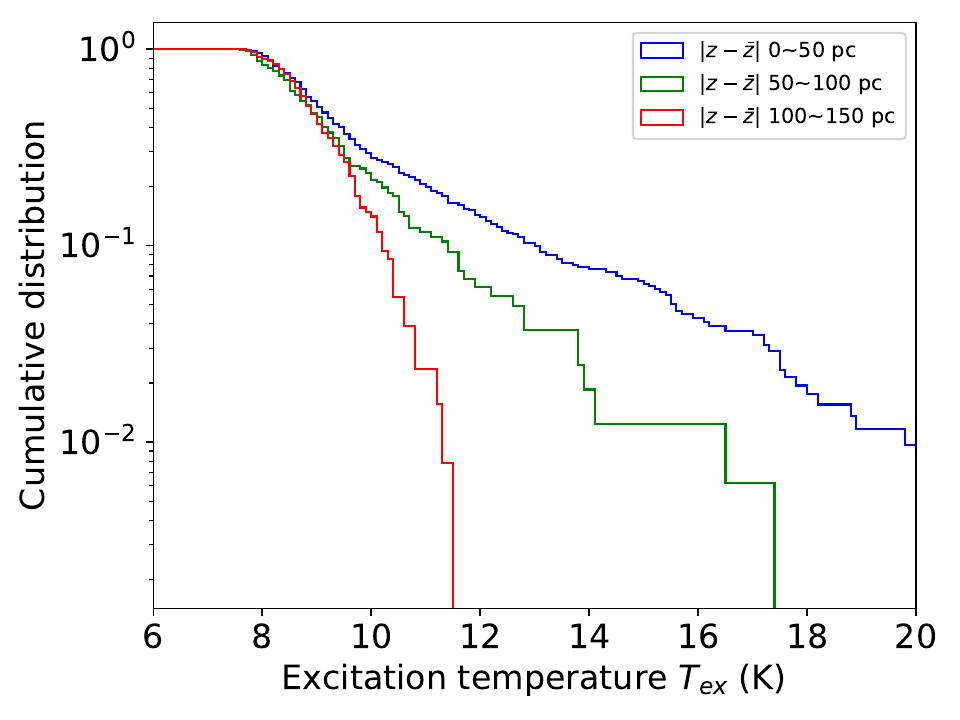}}
    \end{minipage}
    \hfill
    \begin{minipage}{0.48\linewidth}
        \centering
        \subfloat[]{\includegraphics[width=\textwidth]{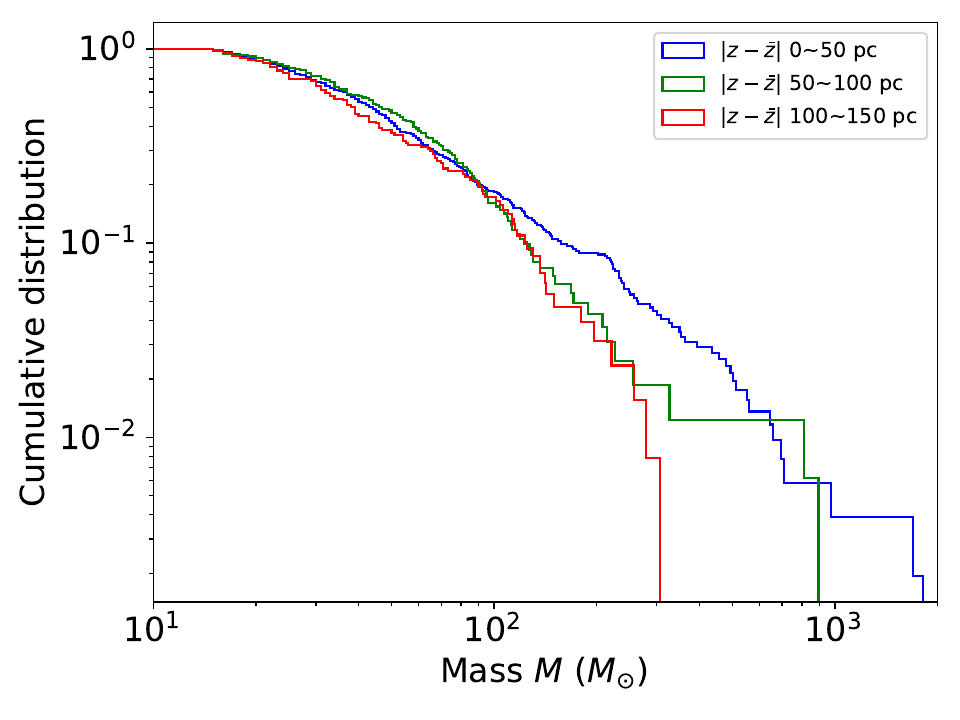}}
    \end{minipage}
        
    \begin{minipage}{0.48\linewidth}
        \centering
        \subfloat[]{\includegraphics[width=\textwidth]{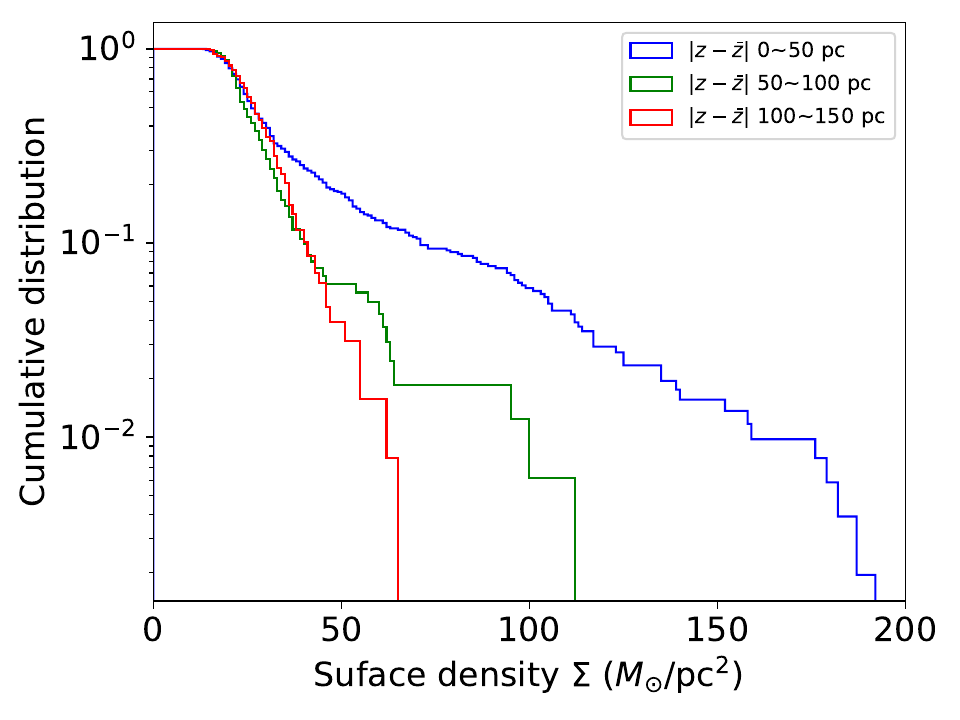}}
    \end{minipage}
    \hfill
    \begin{minipage}{0.48\linewidth}
        \centering
        \subfloat[]{\includegraphics[width=\textwidth]{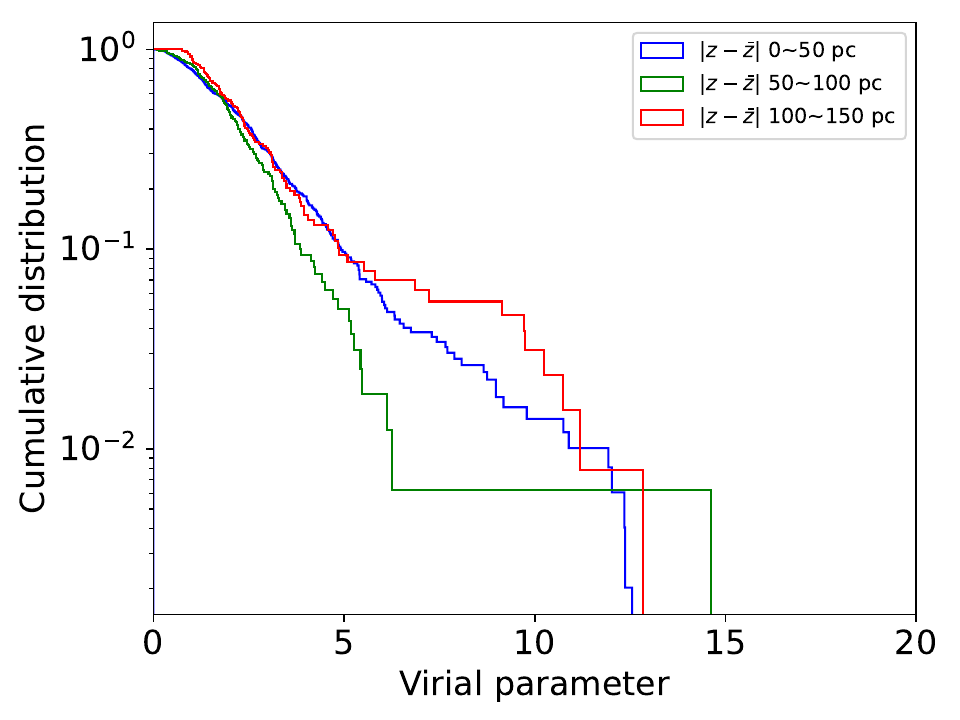}}
    \end{minipage}     
    \caption{Same as Fig. \ref{fig:parameters-G328} but for the G309 region.}
    \label{fig:parameters-G309}
\end{figure*}

The cumulative distributions of the physical parameters for the three sub-groups of the G328 and G309 regions are
shown in Fig. \ref{fig:parameters-G328} and \ref{fig:parameters-G309}, respectively. 

\end{appendix}
\end{document}